\documentclass[preprint2]{proto} 
\usepackage{times} 
\newcommand{\refs}{\par\noindent\hangindent=1pc\hangafter=1} 
\begin{document} 
   
%   
% USEFUL SYMBOLS:    
%   
%\simlt:\la   \simgt:\ga   
%\def\lco{\mbox{$L_{\mbox{\tiny CO}}$}}   
%   
%\def\fco{\mbox{$F_{\mbox{\tiny CO}}$}}   
\def\fco{$F_{\rm CO}$~}   
\def\fcom{F_{\rm CO} }   
\def\lbol{$L_{\rm bol}$~}   
\def\lbolm{L_{\rm bol} }   
\def\tbol{\mbox{$T_{\mbox{\tiny bol}}$}}   
\def\lacc{$L_{\rm acc}$~}   
\def\laccm{L_{\rm acc}}   
\def\k13{\mbox{$\kappa_{\mbox{\tiny 1.3mm}}$}}   
\def\kir{\mbox{$\kappa_{\mbox{\tiny IR}}$}}   
\def\tdust{\mbox{T$_{\mbox{\tiny dust}}$}}   
\def\rout{\mbox{$R_{\mbox{\tiny out}}$}}   
\def\mkmsyr{\rm{M_\odot \, km\, s^{-1}\, {yr}^{-1}}}   
\def\kms{\rm{km\, s^{-1}}}   
\def\myr{\rm{M_\odot\,  yr^{-1}}}   
\def\cmg{\rm{cm^{2} \, g^{-1}}}   
\def\as{\mbox{$a_{\mbox{\tiny s}}$}}   
\def\menv{$M_{\rm env}$~}   
\def\menvm{M_{\rm env}}   
\def\macc{$\dot{M}_{\rm acc}$~}   
\def\maccm{\dot{M}_{\rm acc}}   
\def\mjet{$\dot{M}_{\rm jet}$~}   
\def\mjetm{\dot{M}_{\rm jet}}   
\def\pjet{$\dot{P}_{\rm jet}$~}   
\def\pjetm{\dot{P}_{\rm jet}}   
\def\vjet{$V_{\rm jet}$~}   
\def\vjetm{V_{\rm jet}}   
\def\fent{$f_{\rm ent}$~}   
\def\fentm{f_{\rm ent}}   
\def\nhd{\mbox{$n_{\mbox{\tiny H}_{2}}$}}   
\def\scol{\mbox{$S_{\mbox{\tiny 1.3mm}}^{\mbox{\tiny ~beam}}$}}   
\def\scolnu{\mbox{$S_{\nu}^{\mbox{\tiny ~beam}}$}}   
\def \dss {\displaystyle}   
\def\oneskip{\vskip\baselineskip}   
% other useful symbols:   
\def\>{$>$}   
\def\<{$<$}   
\def\impl{$\Longrightarrow$}   
\def\ltsima{$\; \buildrel < \over \sim \;$}   
\def\simlt{\lower.5ex\hbox{\ltsima}}   
\def\gtsima{$\; \buildrel > \over \sim \;$}   
\def\simgt{\lower.5ex\hbox{\gtsima}}   
\def\arcsec{\hbox{$^{\prime\prime}$}}   
\def\arcmin{$^\prime$}   
\def\beam{\mbox{$_{\mbox{\tiny beam}}$}}   
\def\h2{\mbox{$_{\mbox{\tiny H2}}$}}   
\def\mh{\mbox{m$_{\mbox{\tiny H}}$}}   
\def\K{\mbox{K}}   
\def\mjyb{\mbox{mJy/13\arcsec-beam}}   
\def\nhd{\mbox{$n_{\mbox{\tiny H}_{2}}$}}   
\def\scol{\mbox{$S_{\mbox{\tiny 1.3mm}}^{\mbox{\tiny ~beam}}$}}   
\def\scolnu{\mbox{$S_{\nu}^{\mbox{\tiny ~beam}}$}}   
\def \dss {\displaystyle}   
\def\oneskip{\vskip\baselineskip}   
% other useful symbols:   
\def\>{$>$}   
\def\<{$<$}   
\def\impl{$\Longrightarrow$}   
\def\ltsima{$\; \buildrel < \over \sim \;$}   
\def\simlt{\lower.5ex\hbox{\ltsima}}   
\def\gtsima{$\; \buildrel > \over \sim \;$}   
\def\simgt{\lower.5ex\hbox{\gtsima}}   
\def\arcsec{\hbox{$^{\prime\prime}$}}   
\def\arcmin{$^\prime$}   
\def\beam{\mbox{$_{\mbox{\tiny beam}}$}}   
\def\h2{\mbox{$_{\mbox{\tiny H2}}$}}   
\def\mh{\mbox{m$_{\mbox{\tiny H}}$}}   
\def\K{\mbox{K}}   
\def\mjyb{\mbox{mJy/13\arcsec-beam}}   

\voffset=-0.25in   
   
%Margins:    Inside = 1 in.; outside = 0.75 in.; top = 0.75 in.;   
% bottom = 0.812 in.   
%Columns:  2 columns; space between columns = 0.2 in.   
%Title:  Times Bold, 16/16, centered   
%Author(s):  Times Bold, 12/14, centered   
%Affiliation(s):  Times Bold Italic, 9/11, centered   
%Abstract:  Times, 9/11, full justification; left and right indent 1 in.;   
%first line indent 0.167 in. (1 pic   
%a); auto hyphenation limit = 2; 0.25 in. hyphenation zone   
%Body Text:  Times, 10/12, full justification; first line indent 0.167 in.   
%(1 pica); small   
%caps = 70% of text size; sub and superscript point size = 70% of text point   
%size; sub and superscript position = 25% of text point size; auto hyphen 
%limit = 2; 0.25 in. hyphenation zone   
%Level 1 Headings:  Times Bold, 10/12, centered, all caps   
%Level 2 Headings:  Times Bold, 10/12, left justification, upper and lower  
%Level 3 Headings:  Times Italic, 10/12, full justification   
%Figure Captions:  Times, 9/11, full justification   
%Tables:  Times, 9/11, including table title and tabular data   
%Acknowledgments:  Times, 9/11, full justification   
%References:  Times, 9/11, full justification; left hanging indent 0.167 in.   
%(1 pica)   
   
%\begin{document}   
   
\title{\textbf{\LARGE An Observational Perspective of Low Mass    
Dense Cores II: Evolution towards the Initial Mass Function}}   
   
\author {\textbf{\large Derek Ward-Thompson}}   
\affil{\small\em Cardiff University}   
   
\author {\textbf{\large Philippe Andr\'e}}   
\affil{\small\em Service d'Astrophysique de Saclay}   
   
\author {\textbf{\large Richard Crutcher}}   
\affil{\small\em University of Illinois}   
   
\author {\textbf{\large Doug Johnstone}}   
\affil{\small\em National Research Council of Canada}   
   
\author {\textbf{\large Toshikazu Onishi}}   
\affil{\small\em Nagoya University}   
   
\author {\textbf{\large Christine Wilson}}   
\affil{\small\em McMaster University}

\begin{abstract} 
\baselineskip = 11pt 
\leftskip = 0.65in  
\rightskip = 0.65in 
\parindent=1pc 
{\small   
We review the properties of low mass   
dense molecular cloud cores, including starless,    
prestellar, and Class 0 protostellar cores, as derived from observations.   
In particular we discuss them in the  
context of the current debate surrounding   
the formation and evolution of cores. There exist several families 
of model scenarios to explain this evolution   
(with many variations of each) 
that can be thought of as a continuum of models   
lying between two extreme   
paradigms for the star and core formation process. At one extreme there is   
the dynamic, turbulent picture, while at the other extreme there is a slow,    
quasi-static vision of core evolution. In the latter view  
the magnetic field plays a dominant role,  
and it may also play some role in the former picture. 
Polarization and Zeeman measurements indicate that some, if not all, cores   
contain a significant magnetic field. Wide-field surveys   
constrain the timescales of the core formation and evolution processes,   
as well as the statistical distribution of core masses. The  
former indicates that prestellar cores typically live for 2--5  
free-fall times, while the latter seems to  
determine the stellar initial mass function.  
In addition, multiple surveys allow one to compare core properties  
in different regions. From this it appears that aspects of different  
models may be relevant to different star-forming regions, depending on  
the environment. Prestellar cores in cluster-forming regions are  
smaller in radius and have higher column densities, by up to an order of  
magnitude, than isolated prestellar cores. This is probably  
due to the fact that in cluster-forming regions the prestellar cores are  
formed by fragmentation of larger, 
more turbulent cluster-forming cores, which in turn form as a result 
of strong external compression. It is then the fragmentation  
of the cluster-forming core (or cores) that forms a 
stellar cluster. In more isolated, more quiescent, 
star-forming regions the lower ambient pressure can only support lower   
density cores, which go on to form only a single star or a binary/multiple   
star system. Hence the evolution of cluster-forming cores  
appears to differ from the evolution of more isolated cores.  
Furthermore, for the isolated prestellar cores studied in detail,   
the magnetic field and turbulence appear to be playing a  
roughly equal role.  
 \\~\\~\\~}%leave this in to get the correct vertical space after abstract 
%\end{list}   
\end{abstract}   
   
\section{\textbf{INTRODUCTION}}   
\bigskip   
   
\begin{figure*}   
\setlength{\unitlength}{1mm}  
\noindent  
\begin{picture}(80,100)  
\put(0,0){\includegraphics{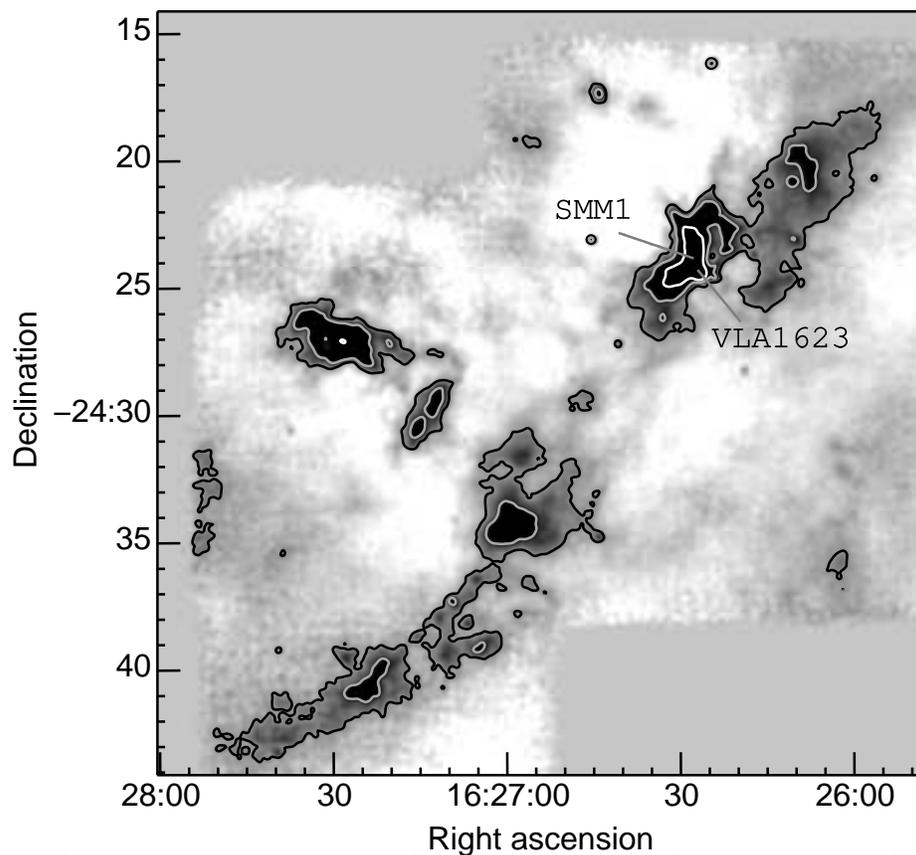}}  
\end{picture}  
\caption{SCUBA image of the $\rho$ Oph molecular cloud region seen   
in dust continuum at 850~$\mu$m (adapted from {\it Johnstone et al.},  
2000). Prestellar, protostellar and cluster-forming    
cores can all be seen in this molecular cloud region. For example,  
the cluster-forming core $\rho$~Oph~A (extended region in the   
upper right of this image) contains within it  
(inside the white contour) the prestellar core  
SM1 and Class~0 
protostellar core VLA1623 (cf. {\it Andr\'e et al.}, 1993). 
Note also that large areas of the cloud contain no dense cores, 
leading to the idea of a threshold criterion discussed in Section 6.}   
\label{rho-oph}   
\end{figure*}   
  
A great deal is now known about dense cores in molecular clouds that   
are the progenitors of protostars -- see the previous chapter   
by {\it Di Francesco et al.}, which details   
many of the observational constraints that have been placed upon their   
physical parameters (this chapter and the preceding chapter should be   
read in conjunction). What is less clear is the manner in which the   
cores are formed and subsequently evolve. In this chapter we discuss    
what the observations can tell us about the formation and evolution    
of cores. Clearly the evolution depends heavily upon   
the formation mechanism, and upon the dominant physics of   
that formation. Several model scenarios have been proposed for this  
mechanism. 
   
These models can be thought of as a small number of families of models, 
each of which contains many variations, 
representing a continuum lying between   
two extremes. At one extreme there is a school of thought that proposes   
a slow, quasi-static evolution, in which a core gradually becomes more    
centrally condensed. This evolution may be moderated by the magnetic field   
(e.g., {\it Mouschovias and Ciolek}, 1999)   
or else by the gradual dissipation of low-level turbulent velocity fields   
(e.g., {\it Myers}, 1998, 2000).   
At the other extreme is a very dynamic picture    
(e.g., {\it Ballesteros-Paredes et al.}, 2003),   
in which highly turbulent   
gas creates large density inhomogeneities, some of which become    
gravitationally unstable and collapse to form stars
(for a review, see: {\it Ward-Thompson}, 2002). Once again the   
magnetic field may play a role in this latter picture, 
in which magneto-hydrodynamic   
(MHD) waves may be responsible for carrying away excess turbulent energy   
(e.g., {\it Ostriker et al.}, 1999).   
   
What we find from the observations is that some aspects of each of these   
different model scenarios may be relevant in different regions of star   
formation, depending on the local environment. No two regions are the same,   
and the effects of local density, pressure and magnetic field strength,    
and the presence or absence   
of other nearby stars and protostars all play  
an important role in determining 
what dominates the formation and evolution of dense molecular cloud cores.   
   
Throughout this chapter we define a dense core as any region in a molecular    
cloud that is observed to be significantly   
over-dense relative to its surroundings. We define a starless    
core as any dense core that does not contain any evidence that it 
harbours a    
protostar, young stellar object or young star ({\it Beichman et al.}, 1986).   
Such evidence would include an embedded infra-red source, centimetre radio    
source or bipolar outflow, for example (cf. {\it Andr\'e et al.}, 1993,   
2000).  
   
Any core that does contain such evidence we define as a protostellar core.   
This might be a Class 0 protostellar core ({\it Andr\'e et al.}, 1993, 2000)   
or a Class I protostellar core ({\it Lada}, 1987; 
{\it Wilking et al.}, 1989)   
depending upon its evolutionary status.   
   
We here define prestellar cores (formerly pre-protostellar cores --   
{\it Ward-Thompson et al.}, 1994) as that subset    
of starless cores which are gravitationally bound and hence are expected to   
participate in the star formation process. 
We further define cluster-forming    
cores as those cores that have significant observed structure within them,    
such that they appear to be forming a small cluster 
or group of stars rather than a    
single star or star system. Examples of  
the various types of cores can be seen in Figure~\ref{rho-oph}.  
   
We note that the resolution of current    
single-dish telescopes is insufficient in more distant regions to    
differentiate between cluster-forming cores and other types of core.   
Hence we restrict most of our discussion to nearby molecular clouds -- 
typically we restrict our discussion to d~$<$~0.5~kpc.   
   
\bigskip   
\centerline{\textbf{ 2. EVOLUTIONARY MODELS}}   
\bigskip   
  
We begin by summarising some of the key model parameters  
and predictions.   
One such discriminator between the extreme pictures mentioned above   
is the timescale of core evolution. Therefore we first discuss  
some predictions of the models regarding core lifetimes.  
  
If turbulent dissipation in a quasi-static scenario   
is the relevant physics, then the timescale of the dissipation of   
turbulence could be several times the free-fall time    
(e.g., {\it Nakano}, 1998). However, if highly turbulent processes   
dominate molecular cloud evolution then   
detailed modelling yields results which suggest that   
cores only live for approximately one or two   
free-fall times (e.g., {\it Ballesteros-Paredes et al.}, 2003; 
{\it Vazquez-Semadeni et al.}, 2005).   
  
In the magnetically-dominated  
paradigm, molecular clouds may form by accumulation of matter along    
flux tubes, by (for example) the Parker instability ({\it Parker}, 1966).  
Furthermore,   
if magnetic fields dominate the evolution then a key parameter is   
the ratio of core mass to magnetic flux ($M/\Phi$).   
A critical cloud or core is defined  
as one in which the energy density of the magnetic field  
exactly balances the gravitational potential energy.  
  
For clouds with magnetic    
fields stronger than is necessary for support  
against gravitational collapse,    
$M/\Phi$ is subcritical; for fields too weak to support clouds, $M/\Phi$ is    
supercritical.  
Consequently, two possible extreme-case scenarios arise: one  
in which low-mass stars form in originally highly magnetically subcritical  
clouds, with ambipolar diffusion leading to core formation and  
quasi-static contraction of the cores (e.g., {\it Mouschovias},  
1991; {\it Shu et al.}, 1987); and the other in which clouds are originally  
supercritical (e.g., {\it Nakano}, 1998).  
In the absence of turbulent support, 
highly supercritical collapse occurs   
on essentially the free-fall time.   
  
Since magnetic fields can be frozen into only the ionized    
component of clouds, neutral matter can be driven by gravity through the    
field. Hence, if a star is formed in  
an originally very magnetically subcritical cloud, the relevant  
timescale is the ambipolar diffusion timescale, $\tau_{AD}$, which  
is proportional to the ionisation fraction $X_e$.  
This is normally taken to   
have a power-law dependence on density: 
$\tau_{AD} \propto X_e \propto n({\rm H}_2)^{-0.5}$ for $A_V > 4$, 
where cosmic-ray ionisation dominates 
({\it McKee}, 1989; {\it Mouschovias}, 1991). 
For $A_V < 4$ UV ionisation dominates, leading to a steeper  
dependence ({\it McKee}, 1989), but this regime 
is not believed to be significant for prestellar cores. 
  
\begin{figure*}   
\setlength{\unitlength}{1mm}  
\noindent  
\begin{picture}(80,80)  
\put(0,0){\includegraphics{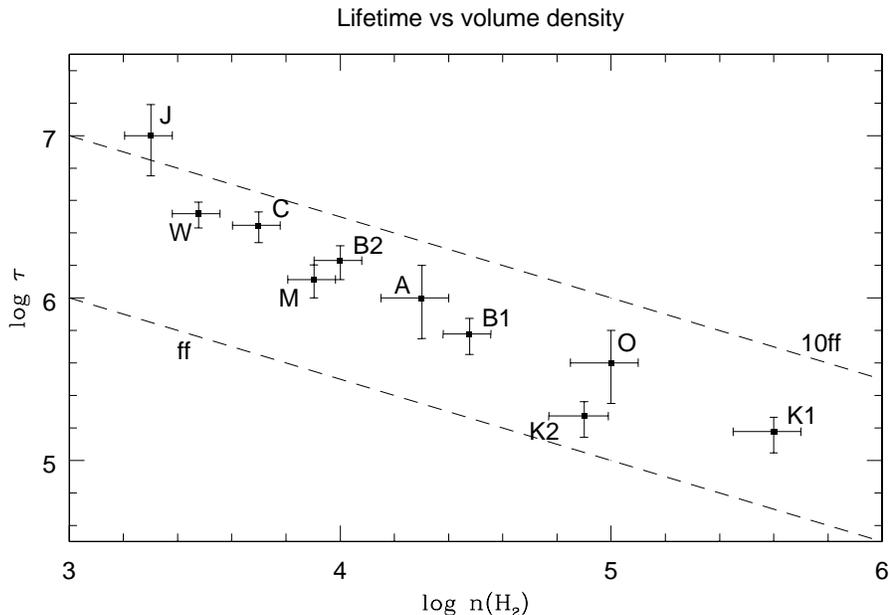}}  
\end{picture}  
\caption{A `JWT plot' (after {\it Jessop and Ward-Thompson}, 2000) -- 
plot of inferred starless core lifetime against mean volume   
density (see also: {\it J. Kirk et al.}, 2005).   
The dashed lines correspond to models discussed in the text.   
The symbols refer to literature data as follows: J -- {\it Jessop and   
Ward-Thompson} (2000); W -- {\it Wood et al.} (1994);    
C -- {\it Clemens and Barvainis} (1988);    
B1, B2 -- {\it Bourke et al.} (1995a, 1995b); M -- {\it Myers et al.}   
(1983); A -- {\it Aikawa et al.} (2005); {\it Kandori et al.} (2005);   
O -- {\it Onishi et al.} (2002); K1, K2 -- {\it J. Kirk et al.} (2005).}   
\label{lifetimes}   
\end{figure*} 
 
Since $\tau_{AD}$ is shorter in denser regions, the process of  
ambipolar diffusion increases $M/\Phi$ in overdense regions  
of the cloud, leading to the formation of cores.  Eventually,  
$M/\Phi$ is increased from subcritical  
to supercritical and the core collapses. For highly  
subcritical clouds $\tau_{AD}$ is roughly ten times the free-fall time  
({\it Nakano}, 1998), although {\it Ciolek and Basu} (2001) point out 
that the ambipolar diffusion timescale of marginally subcritical cores  
within clouds can be as little as a few times the  
free-fall time. Only observations can establish the original  
$M/\Phi$ in clouds; it is a free parameter in the theory.  
   
Magnetic fields may also play another crucial role in star formation --    
transferring angular momentum outward from collapsing, rotating cores,    
resolving the angular momentum problem and allowing 
collapse to protostellar    
densities ({\it Mouschovias}, 1991). 
Although supersonic motions are allowed    
in this paradigm, they do not dominate.   
 
This picture has been challenged by interpretations of observations of the    
ratio of numbers of   
starless cores to cores with protostars and young stars that suggest    
that molecular clouds are short-lived compared with the ambipolar diffusion    
timescale, and that star formation takes place on a cloud-crossing time    
(e.g., {\it Elmegreen}, 2000; {\it Hartmann et al.}, 2001).    
   
This alternative paradigm is that molecular clouds are    
intermittent phenomena in an interstellar medium dominated by compressible    
turbulence (e.g., {\it MacLow and Klessen}, 2004). 
Turbulent flows form density  
enhancements that may or may not be self-gravitating. If self-gravitating,    
they may be supported against collapse for a short time by the turbulent    
energy. However, the supersonic turbulence will 
decay on a short (free-fall)    
timescale (e.g., {\it MacLow et al.}, 1998), and collapse will ensue.    
This would mean that   
molecular clouds were transient objects, forming and either   
dissolving quickly or rapidly collapsing to form stars.     
   
One way to distinguish between the theories is to determine the    
lifetimes of the large-scale molecular clouds.    
{\it Hartmann et al.} (2001) argue for short lifetimes, whereas   
{\it Tassis and Mouschovias} (2004) and {\it Mouschovias et al.} (2006) 
suggest that    
all available observational data are consistent with lifetimes of molecular    
clouds as a whole being $\sim 10^7$ yr (see also {\it Goldsmith and Li}, 
2005). Another way is to determine the lifetimes of individual cloud cores.  
We attempt to do this in the next section. The role of magnetic fields  
can be assessed by measuring the $M/\Phi$ values in cores. We discuss  
the current data on this in Section 4.  
   
\bigskip  
\centerline{\textbf{ 3. OBSERVED CORE LIFETIMES}}  
\bigskip  
  
It was shown in the previous section that   
it is of vital importance to estimate observationally   
the timescale of cores with various densities if we are to distinguish   
between the different model pictures.   
The numbers of cores detected can be used to determine typical statistical   
timescales for particular evolutionary stages. This method was first   
used by {\it Beichman et al.} (1986),    
who extrapolated from the typical T Tauri    
star lifetime and estimated the starless core   
lifetime to be roughly a few times 10$^6$ years.   
   
This estimate was subsequently refined by {\it Lee and Myers} (1999), using    
an optically-selected sample, to $\sim$0.3--1.6 $\times$ 10$^6$ years 
for a mean density of $\sim$6--8 $\times$ 10$^3$ cm$^{-3}$. 
This age is based upon an estimated range in lifetimes for Class I sources   
of $\sim$1--5 $\times$ 10$^{5}$ years. Within this range the best estimate   
for the Class I lifetime is $\sim$2 $\pm$ 1 $\times$ 10$^5$ years   
(e.g., {\it Greene et al.}, 1994; {\it Kenyon and Hartmann}, 1995).   
This corresponds to a starless core lifetime of $\sim$6 $\pm$ 3 $\times$   
10$^5$ years.   
   
Towards some molecular cloud complexes optical   
selection can miss deeply embedded cores in the complex.  In these cases,   
observations in the mm/submm regime are the best way to observe cores   
and to carry out the statistical study. For example,   
{\it Onishi et al.} (1998, 2002) estimated the time-scale of cores    
with a density of $\sim$10$^5$ cm$^{-3}$ to be   
$\sim$4$\times$10$^5$ years,   
based on a large-scale molecular line study of cores in Taurus.   
   
{\it J. Kirk et al.} (2005) carried out a similar exercise using submm   
continuum observations of dust in molecular cloud cores. They found a    
timescale for pre-stellar cores of $\sim$3 $\times$ 10$^5$ years with a   
minimum central density of   
$\sim$5 $\times$ 10$^4$ cm$^{-3}$. At this density the free-fall time   
is $\sim$10$^5$ years.    
They made a similar calculation for the cores they classified as `bright',   
and derived a time-scale of $\sim$1.5 $\times$ 10$^5$ years for cores with a   
minimum central volume density of $\sim$2 $\times$ 10$^5$ cm$^{-3}$.   
At this density the free-fall timescale is $\sim$7$\times$ 10$^4$ years.   
  
{\it Kandori et al.} (2005) derived detailed radial column density profiles   
for Bok globules and, by comparison with the theoretical calculations of   
{\it Aikawa et al.} (2005), estimated their timescale to be   
$\sim$10$^6$ years for a density of $\sim$2$\times$10$^4$ cm$^{-3}$.   
A number of other chemical models have been used to carry out a similar   
exercise in estimating the `chemical age' of cores (see previous chapter   
by {\it Di Francesco et al.}).   
In many cases this leads to values much longer than a free-fall time.   
   
A similar comparison to that discussed in this section  
was carried out for a number of different data-sets    
in the literature by {\it Jessop and Ward-Thompson} (2000), who plotted the    
calculated statistical lifetime against the mean volume density of each    
sample of cores. We reproduce those data here in Figure~\ref{lifetimes},    
along with other, more recent data. For example, we include the   
`bright' and `intermediate' cores from {\it J. Kirk et al.} (2005) --   
labelled K1 and K2 respectively.   
   
\begin{figure*}   
\setlength{\unitlength}{1mm}  
\noindent  
\begin{picture}(80,80)  
\put(0,0){\includegraphics{wardthompson_fig3a.ps}}   
\put(0,0){\includegraphics{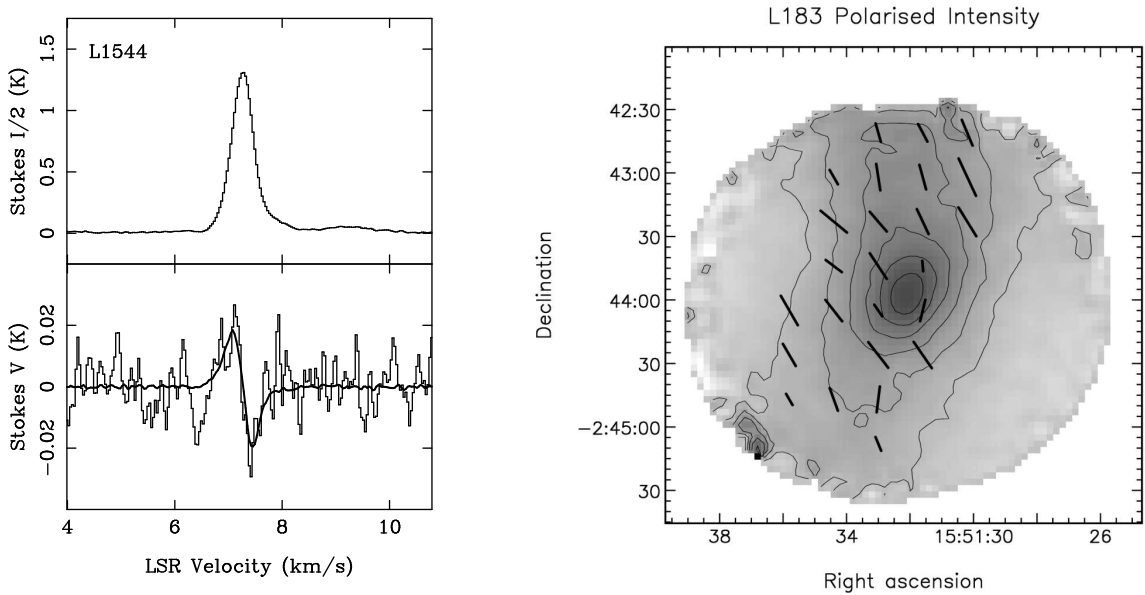}}  
\end{picture}  
\caption{Plot of Arecibo OH Zeeman spectra (left) in the  
prestellar core L1544, from {\it Crutcher and Troland} (2000), and 
SCUBA submillimetre dust intensity and  
polarised intensity data of prestellar core L183 (right),  
from {\it Crutcher et al.} (2004). 
The polarised intensity half   
vectors have been rotated by 90$^\circ$ to show 
the plane-of-sky magnetic field direction.}   
\label{L1544}   
\end{figure*}   
  
We also plot on Figure~\ref{lifetimes}   
some of the model predictions discussed above (following  
{\it J. Kirk et al.} 2005). 
The lower dashed line is the free-fall time, relevant to models   
such as the highly magnetically supercritical models and the   
highly dynamic, turbulent models 
(e.g., {\it Vazquez-Semadeni et al.}, 2005).   
The upper dashed line is the power-law formulation of  
{\it Mouschovias} (1991) discussed in Section 2 above,  
for a quasi-static, magnetically subcritical core evolving  
on the ambipolar diffusion timescale at  
ten times the free-fall time ({\it Nakano}, 1998).  
  
In summary, almost all of the literature estimates  
lie between the two dashed lines on Figure~\ref{lifetimes}.   
All of the observed timescales are longer than the free-fall time by a   
factor of $\sim$2--5 in the density range of 10$^4$--10$^5$ cm$^{-3}$.     
Hence we see quite clearly that prestellar and starless cores  
cannot generally   
all be in free-fall collapse. Their timescales also appear to be too short   
for them all to be in a highly magnetically subcritical state.    
They are all roughly consistent both with mildly subcritical magnetised 
cores and with models invoking low levels of turbulent support. 
Hence we must look to observations of magnetic fields to help differentiate 
between models. In the next section we summarise some of the key observations 
of magnetic fields. 
  
\bigskip   
\centerline{\textbf{ 4. OBSERVATIONS OF MAGNETIC FIELDS}}   
\bigskip   
   
Given that the relative importance of the magnetic field is a key  
way in which to choose between models, one  
must try to determine observationally  
the role of magnetic fields in the star formation process.   
   
Many  
observations of magnetic fields in regions of low-mass star formation have    
attempted to test the various paradigms. The observations have    
utilized the Zeeman effect, mainly in the 18-cm lines of OH  
(e.g., {\it Crutcher et al.}, 1993; {\it Crutcher and Troland}, 2000),    
and linearly polarized emission    
of dust at submillimetre wavelengths 
(e.g., {\it Ward-Thompson et al.}, 2000;    
{\it Matthews et al.}, 2001; {\it Crutcher et al.}, 2004;   
{\it J. Kirk et al.}, 2006).    
   
Unfortunately, the observations are difficult and    
the results remain somewhat sparse (see, e.g., {\it Crutcher}, 1999; 
{\it Heiles and Crutcher}, 2005).    
One of the best-studied prestellar  
cores is L1544, a relatively isolated  
core in Taurus that has been studied by    
single-dish ({\it Tafalla et al.}, 1998) and interferometer spectroscopy    
({\it Williams et al.}, 1999). These studies have suggested that L1544 is    
contracting.    
   
Information on the magnetic field in L1544 includes OH Zeeman    
observations with the Arecibo telescope ({\it  Crutcher and Troland}, 2000)    
and dust polarization mapping with the JCMT SCUBA polarimeter    
({\it Ward-Thompson et al.}, 2000). This cloud is therefore a good example    
of observational results in low-mass star formation regions.   
   
Figure~\ref{L1544} shows the Arecibo OH Zeeman spectra, which imply a   
line-of-sight (los) magnetic field strength of $B_{los}    
= +10.8 \pm 1.7$ $\mu G$, with column density   
$N(H_2) \approx 4.8 \times 10^{21}$ cm$^{-2}$, mean radius    
$\overline{r}(OH) \approx 0.08$ pc, and volume density   
$n(H_2) \approx 1 \times 10^4$ cm$^{-3}$.  
Because all three components of the magnetic field vector are not 
generally observed, and because the inferred column densities are not 
generally along the direction of the magnetic field vector, the 
directly observed $M/\Phi$ is typically an overestimate of the true value. 
A statistical correction for this is possible (see {\it Heiles and  
Crutcher}, 2005). For a large randomly oriented sample, the observed  
$M/\Phi$ average should be divided by 3 to obtain the statistically  
correct result. This correction may be applied to each cloud individually, 
but it must be kept in mind that this correction is only strictly 
valid for a large sample of measurements. 
The directly observed $M/\Phi$ for L1544 is $\approx 3.4$. 
{\it Crutcher and Troland} (2000) corrected this value    
statistically for geometrical bias, finding $M/\Phi \approx 1.1$, or   
roughly critical. 
   
Figure~\ref{L1544} also shows the SCUBA dust intensity and polarized  
intensity map of L183 ({\it Crutcher et al.}, 2004).  
{\it Ward-Thompson et al.} (2000) had previously mapped L183 and  
two other prestellar cores -- L1544 and L43.   
{\it Crutcher et al.} (2004) used the   
Chandrasekhar-Fermi (CF)    
method ({\it Chandrasekhar and Fermi}, 1953) to measure  
the magnetic field strengths and hence the relative criticality  
of all three cores.  
  
In Figure~\ref{L1544}   
the polarisation half vectors have been rotated by 90$^\circ$ to    
indicate the plane-of-sky (pos) magnetic field direction (a half vector is   
a vector with a 180$^\circ$ bi-directional ambiguity, such as we have here).   
The field is seen to be fairly uniform in direction  
in L183, as it was in L1544 and L43, with a position angle    
dispersion $\delta \phi \approx 14^\circ$, but the direction of the field    
in the plane of the sky is at an angle of $34^\circ \pm 7^\circ$ to the    
minor axis.    
   
This difference between the magnetic field direction and the    
minor axis of the core is in conflict with symmetric models that rely only    
on thermal and static magnetic pressure to balance gravity, since the minor    
axis projected onto the sky should lie along $B_{pos}$. However, projection    
effects can produce the observed position angle difference if the core has    
a more complicated shape, such as a triaxial geometry ({\it Basu}, 2000).    
The initial    
conditions of cloud formation and turbulence may produce the more    
complicated shapes ({\it Gammie et al.}, 2003).    
   
The physical parameters of the L1544 prestellar  
core inferred from the SCUBA data ({\it Crutcher et al.}, 2004) are:    
$\overline{r}(dust) \approx 0.021$ pc, $N(H_2) \approx 4.2 \times 10^{22}$    
cm$^{-2}$, $n(H_2) \approx 4.9 \times 10^5$ cm$^{-3}$, and total mass   
$M \approx 1.3$    
$M_\odot$.  With the velocity dispersion   
$\Delta$V$_{NT} \approx 0.28$ km s$^{-1}$, as measured from N$_2$H$^+$    
data ({\it Caselli et al.}, 2002), the CF method   
yielded $B_{pos} \approx 140$ $\mu$G. Then $M/\Phi \approx 2.3$, 
({\it Crutcher et al.}, 2004) and 
the statistically corrected value is then $M/\Phi \approx 0.8$. Hence L1544 
is approximately critical or mildly supercritical. 
 
\begin{figure*}   
\setlength{\unitlength}{1mm}  
\noindent  
\begin{picture}(80,75)  
\put(0,0){\includegraphics{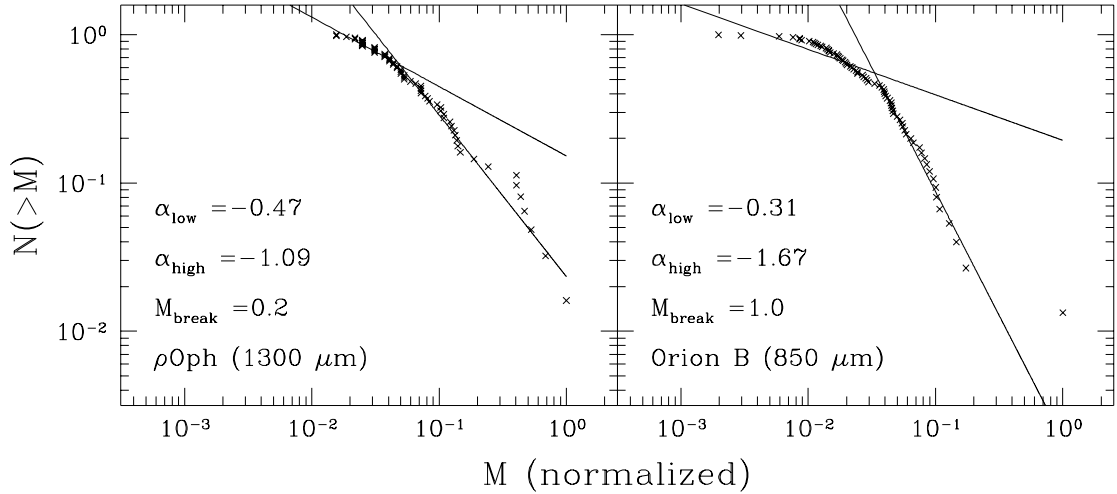}}  
\end{picture}  
\caption{Plot of normalised 
core mass function for the $\rho$ Oph and Orion  
molecular clouds (adapted from {\it Reid}, 2005; based on  
original data from {\it Motte et al.}, 1998; and  
{\it Johnstone et al.}, 2000). The two slopes  
and break-point mass of double power law fits are given in each panel.  
The break-point masses are quoted in M$_\odot$. 
Note the similarity to the stellar initial mass function  
(e.g., {\it Kroupa}, 2002).}   
\label{cmf}   
\end{figure*}  
 
The $M/\Phi$s for L1544   
found from the OH Zeeman and the dust polarization techniques are    
essentially in agreement, but very different regions are sampled by the two    
methods. The region sampled by OH has 4 times the radius, 0.1   
times the column density, and 0.02 times    
the volume density of the region sampled by the dust    
emission. Therefore, the data probe separately the envelope and the core    
regions of the cloud.    
   
We have argued that the two $M/\Phi$ values are consistent to within   
the errors, but if the difference between them were real, then $M/\Phi$    
decreases from envelope to core, the opposite of the ambipolar diffusion    
prediction. However, the Zeeman effect measures $B_{los}$ and dust emission    
measures $B_{pos}$, and we do not know the inclination of \textbf{B} to the    
line of sight. Direct measurement of an increase in $M/\Phi$ from    
envelope to core would strongly support the ambipolar diffusion model.    
However, present data do not allow one to do this.   
   
Other prestellar cores have recently been mapped in submm polarisation.    
{\it J. Kirk et al.} (2006) mapped two cores, L1498 and L1517B.  
They measured 
the magnetic field strength by the CF method and estimated both the   
(non-magnetic) virial mass and the magnetic critical mass. 
In both cases they   
found the prestellar cores to be super-critical by a factor of $\sim$2--3. 
For comparison the three   
cores of {\it Crutcher et al.} (2004) were also seen to be mildly    
supercritical, as predicted by the ambipolar diffusion model. 
   
However, when {\it J. Kirk et al.} (2006) calculated the magnetic virial   
mass (i.e. including the effects of both magnetic fields and turbulent   
line-widths) they found the cores to be roughly virialised, with the    
magnetic field providing roughly half of the support (as was the case   
for the three cores studied by {\it Crutcher et al.} 2004).   
   
Hence we see that for the five prestellar cores whose magnetic fields   
have been studied in detail,   
both turbulence and magnetic fields are seen to be playing a roughly   
equal role in the support against gravitational collapse, and thus in the 
evolution of the cores ({\it J. Kirk et al.}, 2006).  
These cores are all relatively isolated   
and moderately quiescent cases,   
and all give a similar result. Thus we may perhaps conclude 
that for isolated   
star formation in fairly quiescent molecular clouds, one must consider the   
influences of mildly turbulent motions and magnetic fields together.   
Finally, we note that similar submillimetre dust polarization results 
have been obtained for a number of Class 0 protostellar cores 
(e.g., {\it Matthews and Wilson}, 2002; {\it Wolf et al.}, 2003). 
   
\bigskip   
\centerline{\textbf{ 5. CORE MASS FUNCTION}}   
\bigskip   
   
The last seven years since PPIV have seen some progress in measuring the    
mass function of cold cores in molecular clouds with a wide range of   
intrinsic mass scales. This progress has been made possible by the   
availability of new, sensitive cameras at millimetre and submillimetre   
wavelengths (e.g., {\it Kreysa et al.}, 1999; {\it Holland et al.}, 1999).    
   
\begin{figure*}   
\setlength{\unitlength}{1mm}  
\noindent  
\begin{picture}(80,90)  
\put(0,0){\includegraphics{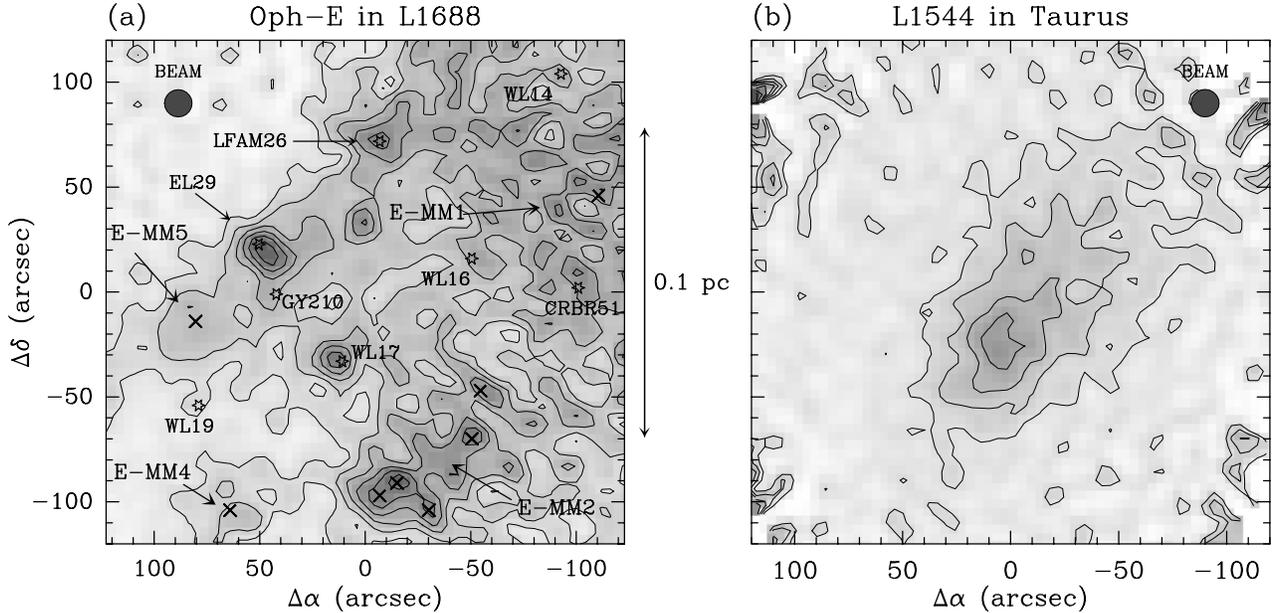}}  
\end{picture}  
\caption{Millimetre dust continuum images of $\rho$ Oph-E (left) and Taurus    
(right) taken at the same resolution with the IRAM 30-m telescope (from 
{\it Motte et al.}, 1998 and {\it Ward-Thompson et al.}, 1999 respectively). 
Note how the cluster-forming   
$\rho$~Oph core shows far more substructure than the more isolated Taurus   
pre-stellar core at the same linear scale. 
In the left-hand image the crosses
mark starless cores and the stars mark protostellar objects.}   
\label{clusterisol}   
\end{figure*}   
   
Wide-area millimetre continuum 
mapping of the Ophiuchus molecular cloud ({\it Motte et al.}, 1998)   
first revealed a core mass function   
that bears a striking similarity to the stellar initial mass function   
(IMF -- e.g., {\it Kroupa}, 2002; {\it Chabrier}, 2003).    
This was subsequently confirmed by others -- e.g., in   
Serpens ({\it Testi and Sargent}, 1998),   
$\rho$ Oph ({\it Johnstone et al.}, 2000;   
{\it Reid and Wilson}, 2006; {\it Stanke et al.}, 2006), and Orion   
({\it Motte et al.} 2001; {\it Johnstone et al.}, 2001, 2006;   
{\it Nutter}, 2004; {\it Reid and Wilson}, 2006).   
Figure~\ref{cmf} shows the core mass functions for $\rho$~Oph   
and Orion ({\it Reid and Wilson}, 2006), based on the 
original results of {\it Motte et al.} (1998) and {\it Johnstone et al.} 
(2000). 
   
In all of these regions the slope of the cumulative 
core mass function above 0.5--1 M$_\odot$    
is $-$1.0 to $-$1.5 (see Figure~\ref{cmf}), in good agreement with   
the high-mass slope of $-$1.35 for the stellar IMF ({\it Salpeter}, 1955).    
The core mass function is observed to have a shallower slope   
at smaller masses, although it has been questioned as to   
whether the change in slope at lower masses is an intrinsic property   
of the clump mass function or is caused by some kind of incompleteness   
in the observations (e.g., {\it Johnstone et al.}, 2000, 2001).   
   
In addition, the peak of the core mass function    
in each of these cluster-forming regions   
lies in the range of 0.2--1 M$_\odot$ (in $dN/d{\rm log}M$ format),  
only slightly larger than   
the peaks of the mass functions of    
$\sim 0.08$ M$_\odot$  for single stars  and $\sim 0.2$ M$_\odot$   
for multiple systems ({\it Chabrier}, 2003).   
In short, both the shape and the intrinsic scale of the core mass   
function in these regions appear to be well-matched to the observed   
properties of the stellar IMF.   
   
Some similar work on more distant, higher mass regions has also been carried   
out (e.g., {\it Tothill et al.}, 2002;    
{\it Motte et al.}, 2003; {\it Mookerjea et al.}, 2004;   
{\it Beuther and Schilke}, 2004; {\it Reid and Wilson}, 2005, 2006),    
although this is   
strictly beyond the scope of this chapter on low-mass cores. In addition,   
these studies suffer from problems such as: a cluster of low-mass   
cores can appear, in these more distant regions,   
to merge into a single higher-mass core; any incompleteness in   
the mass function will set in at relatively higher masses;   
and most of these studies make no distinction between starless cores and    
those with protostars.   
   
It has been suggested that   
the fact that the shape of the core mass function does not appear to   
vary from region to region even as its intrinsic scale is   
changing, appears to be consistent with the core mass function being   
determined primarily by turbulent fragmentation (e.g., {\it Reid}, 2005).   
However, this result is subject to the caveats mentioned above.   
Nonetheless, numerical simulations by several groups have shown that   
turbulent fragmentation can produce clump mass functions whose shape   
does not depend strongly on the intrinsic mass scale of the region   
({\it Klessen et al.}, 1998; {\it Klessen and Burkert}, 2000; 
{\it Klessen}, 2001a; 
{\it Padoan and Nordlund}, 2002; {\it Gammie et al.}, 2003;    
{\it Tilley and Pudritz}; 2004).   
   
\bigskip   
\centerline{\textbf{ 6. CLUSTER-FORMING VS ISOLATED CORES}}   
\bigskip   
 
The environment in which a core forms is crucial to its subsequent 
evolution. This has been known for some time. The sequential model 
of star formation ({\it Lada}, 1987) predicts that where young stars 
have already formed, their combined effects will cause further star 
formation in the remainder of the molecular cloud. This was seen, 
for example, in the $\rho$ Oph molecular cloud, where {\it Loren} 
(1989) hypothesised that the upper Sco OB association was triggering 
star formation in L1688. Further evidence in support of this  
hypothesis was provided by a comparison of the 
relative star-formation activity in L1688 and L1689 ({\it Nutter et al.}, 
2006), wherein these two adjacent clouds were seen to have very different  
levels of star formation due to L1689 being further from the OB association. 
 
Furthermore, the dense cores that are observed on a $\sim 0.1$~pc scale    
in nearby cluster-forming clouds using classical high-density tracers,    
such as NH$_3$, N$_2$H$^+$, H$^{13}$CO$^+$, DCO$^+$, C$^{18}$O,    
and dust continuum emission, tend to have higher masses and    
column densities than isolated prestellar cores    
(e.g., {\it Jijina et al.}, 1999).   
Figure~\ref{clusterisol} shows a comparison between the Taurus and 
central Ophiuchus   
star-forming regions. It can be seen that the region occupied by   
a typical single prestellar core in Taurus    
plays host to a small cluster in Ophiuchus.   
Moreover, the level of cluster-forming activity in a core    
clearly correlates with core mass and column density    
(e.g., {\it Aoyama et al.}, 2001).    
 
\begin{figure*}   
\setlength{\unitlength}{1mm}  
\noindent  
\begin{picture}(80,75)  
\put(0,0){\includegraphics{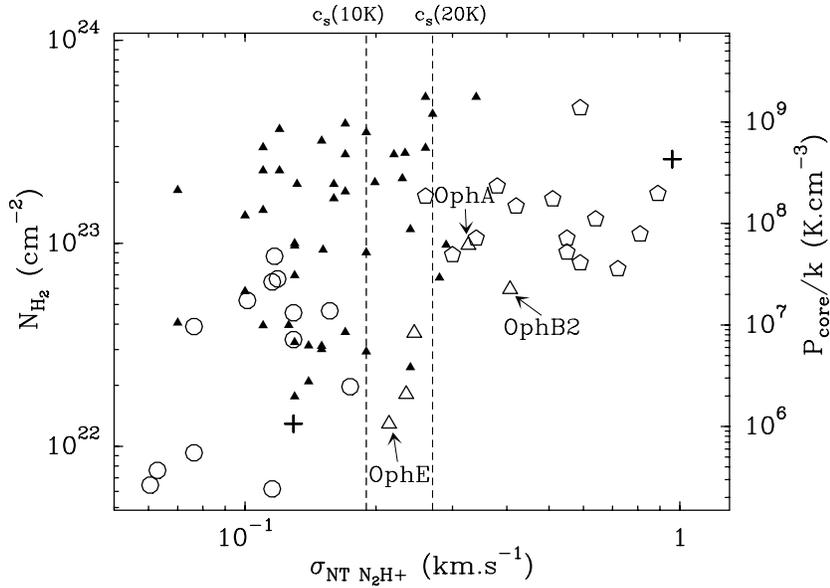}}  
\end{picture}  
\caption{Plot of column density against non-thermal velocity dispersion  
in low-mass dense cores (following {\it Andr\'e et al.}, 2006). 
The two vertical dashed lines represent the sound 
speeds for 10K and 20K gas respectively. The open circles represent the  
isolated prestellar cores in Taurus. The large, open triangles are the   
cluster-forming cores in $\rho$~Oph. The small, filled triangles are the   
prestellar cores in $\rho$~Oph ({\it Motte et al.}, 1998;  
{\it Andr\'e et al.}, 2006). The large, open pentagons are the    
cluster-forming cores in NGC2264D ({\it Peretto et al.}, 2006). 
The two crosses are two equilibrium models 
representative of quasi-static scenarios for low-mass (lower left) 
and high-mass (upper right) star formation (cf.  
{\it Shu et al.}, 1987; and {\it McKee and Tan}, 2003, respectively). 
Note the broad trend seen in the open symbols, and that  
cluster-forming cores lie on the right-hand (supersonic) side 
of the figure, while prestellar cores lie on the left-hand (subsonic) 
side. Note also that most of the prestellar cores in the    
cluster-forming region of $\rho$~Oph lie above the `sequence' observed in    
the open polygons, at up to an order of magnitude higher densities 
than isolated prestellar cores.}   
\label{nh2vnt}   
\end{figure*}   
 
High column-density, cluster-forming cores are typically fragmented    
and show a great deal of substructure (see Figure~\ref{clusterisol}a).   
Submillimetre dust continuum mapping of    
the $\rho$~Oph, Serpens, and Orion~B cluster-forming cores    
has revealed a wealth of compact starless and pre-stellar cores    
(e.g., {\it Motte et al.}, 1998, 2001; {\it Johnstone et al.}, 2000,    
2001; {\it Kaas et al.}, 2004; {\it Testi and Sargent}, 1998), which appear   
to be the direct precursors of individual stars or systems. In particular,    
their mass distribution is remarkably similar to the stellar IMF    
(see Section~5 above).    
   
These prestellar cores in clusters are denser    
($<n> \simgt 10^6$--$10^7$~cm$^{-3}$), more compact    
(diameter $D \sim $~0.02--0.03~pc), and more closely spaced    
($L \sim $~0.03~pc) than isolated prestellar cores,   
such as those seen in Taurus, which typically have     
$<n> \simgt 10^5$~cm$^{-3}$, $D \sim $~0.1~pc, and $L \sim $~0.25~pc.   
(e.g., {\it Onishi et al.}, 2002; previous chapter by  
{\it Di Francesco et al.}).    
   
We define the local star-forming efficiency (SFE$_{pre}$) associated with 
a prestellar core as:   
   
\[ SFE_{pre} = \frac{M_{*}}{M_{pre}} \]  
   
\noindent   
where $M_{pre}$ is the initial 
mass of a prestellar core that forms a star of mass   
$M_{*}$. We find that   
the star formation efficiency within prestellar cores in   
cluster-forming regions is high.   
Most of their initial mass at the onset of collapse    
appears to end up in a star or stellar system: 
$SFE_{pre} \geq$ 50\% 
(cf. {\it Motte et al.}, 1998;    
{\it Bontemps et al.}, 2001). This contrasts with the lower ($\sim 15\% $)    
local star formation efficiency associated with the isolated prestellar  
cores in Taurus ({\it Onishi et al.}, 2002).   
   
Interestingly, extensive searches for cores    
in the Ophiuchus and Pipe Nebula complexes 
(e.g., {\it Onishi et al.}, 1999;    
{\it Tachihara et al.}, 2000; {\it Johnstone et al.}, 2004;    
{\it Nutter et al.}, 2006) suggest     
that cluster-forming cores and prestellar cores can only form    
in a very small fraction of the volume of any given molecular 
cloud complex,    
typically at a compressed extremity.   
  
Recent analysis of the physical conditions within cluster-forming   
molecular clouds has revealed an apparent extinction    
threshold criterion. {\it Johnstone et al.} (2004) noted that   
in Ophiuchus almost all cores were located in high   
extinction regions ($A_V > 10$), despite the fact that most of the cloud   
mass was found at much lower extinctions (cf. Figure~1). 
   
Analysis of the Perseus cloud ({\it H. Kirk}, 2005;  
{\it H. Kirk et al.}, 2006; 
{\it Enoch et al.}, 2006) reveals a similar extinction threshold,   
although at a somewhat lower value ($A_V > 5$). These results are in   
agreement with the analysis of Taurus using C$^{18}$O by    
{\it Onishi et al.} (1998),    
who found that only regions with column densities above   
N(H$_2$)$ = 8 \times 10^{21}\,$cm$^{-2}$ ($A_v \sim 8$) contained IRAS   
sources, indicating that high column density is necessary 
for star formation.   
   
These observations are in fact consistent with the idea that    
magnetic fields play an important   
role in supporting molecular clouds    
({\it McKee}, 1989). The outer region of the   
cloud is maintained at a higher fractional ionization level by   
ultraviolet photons from the interstellar radiation field. However, the    
ultraviolet photons cannot penetrate deep into the cloud   
due to extinction.   
   
The ionization fraction thus drops in the inner region, shortening    
the ambipolar diffusion timescale. According to this scenario, one    
expects small-scale structure and star formation to proceed only in the    
inner, denser regions of the cloud. McKee estimates the column   
density depth required for sufficient ultraviolet attenuation to be   
in the region of $A_V \sim 4-8$. 
 
Column density and linewidth are among the 
key parameters for a core in determining   
its evolution. Figure~\ref{nh2vnt} plots  
the column density N(H$_2$) versus the 
non-thermal velocity dispersion $\sigma_{NT}$, measured in the N$_2$H$^+$   
line, for a large number of cores. The open circles are isolated prestellar   
cores. The open triangles and pentagons are cluster-forming cores.  
The filled triangles are prestellar cores in cluster-forming regions.   
   
It is seen that the cluster-forming cores of, for example, Ophiuchus, 
Serpens, Perseus and Orion have linewidths dominated by non-thermal    
motions (e.g., {\it Jijina et al.}, 1999; {\it Aso et al.}, 2000).     
They are significantly more turbulent    
than the more isolated prestellar cores of Taurus, whose linewidths are    
dominated by thermal motions (e.g., {\it Tatematsu et al.}, 2004;   
{\it Benson and Myers}, 1989). In plotting Figure~\ref{nh2vnt} we have   
removed the thermal velocities and plot only the non-thermal velocity   
dispersions.   
   
However, the more compact ($\sim 0.03$~pc) prestellar cores    
observed within cluster-forming regions are characterized    
by fairly narrow N$_2$H$^+$(1-0) linewidths    
($\Delta V_{FWHM} \simlt 0.5$ km~s$^{-1}$),    
more reminiscent of the isolated prestellar cores of Taurus    
({\it Belloche et al.}, 2001).   
This indicates subsonic or at most transonic    
levels of internal turbulence   
and suggests that, even in cluster-forming clouds,    
the initial conditions for individual    
protostellar collapse are relatively   
free of supersonic turbulence.      
It can be seen from Figure~\ref{nh2vnt} that   
the nonthermal velocity dispersion,    
measured toward the prestellar cores of the $\rho$~Oph cluster   
is only a fraction (typically 0.5--1) of the isothermal sound speed   
({\it Belloche et al.}, 2001; {\it Andr\'e et al.}, 2006).     
   
The narrow N$_2$H$^+$ line widths measured    
for the prestellar cores in these clusters imply virial masses     
which generally agree well with the mass estimates derived from    
the dust continuum. This confirms that most of the starless cores    
identified in the submm dust continuum    
are self-gravitating and very likely prestellar in nature.    
  
Furthermore, Figure~\ref{nh2vnt} appears to show that the prestellar   
cores in a cluster-forming region such as $\rho$~Oph (filled triangles)   
largely occupy a different parameter space from the isolated prestellar   
cores of Taurus (open circles). This tends to imply a different   
formation mechanism 
for prestellar cores in isolated and clustered regions, with   
the latter forming by fragmentation of higher-mass, more turbulent, 
cluster-forming cores (cf. {\it Myers}, 1998).   
For this reason (following {\it Motte et al.}, 2001), 
we suggest that prestellar cores in clustered regions could 
perhaps be called prestellar condensations to indicate this difference. 
 
In addition, there appears to be a broad trend of increasing velocity 
dispersion with increasing column density from isolated prestellar cores 
to cluster-forming cores (cf. {\it Larson}, 1981). 
This perhaps reflects the fact that all of these cores 
are self-gravitating, hence characterized by virial mass ratios close 
to unity. Higher density cores form in clustered regions,  
where the ambient pressure is higher, 
and subsequently fragment before forming stars.   
 
\begin{figure*}   
\epsscale{1.5}   
\plotone{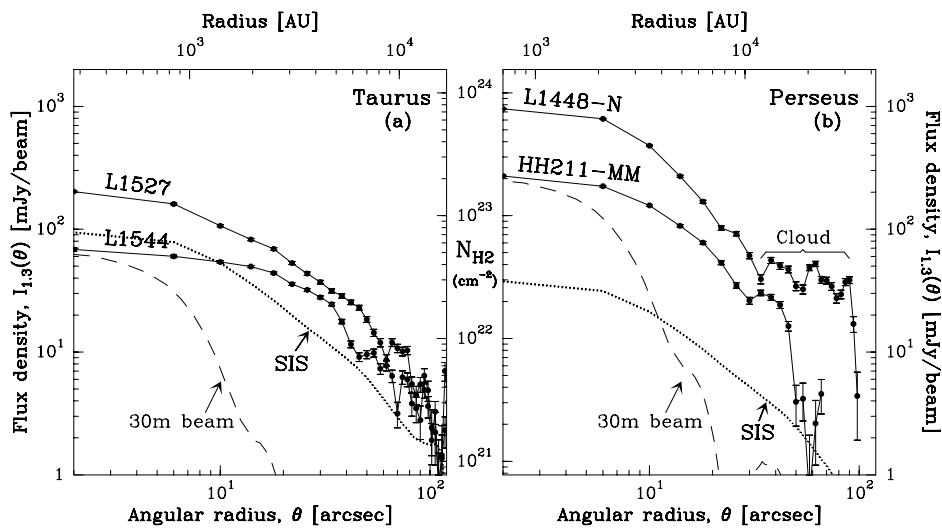}   
\caption{Radial density profiles in Taurus (left)  
and Perseus (right) -- from {\it Motte and Andr\'e} 2001. 
In Taurus we show a prestellar core, L1544,  
and a Class 0 protostar, L1527.  
In Perseus we show two Class 0 protostars, L1448-N and HH211-MM. 
The dotted line marked 
SIS shows the initial conditions for spontaneous collapse  
(e.g., {\it Shu et al.}, 1987) convolved with the beam (dashed line). 
Note that the prestellar core has a flatter profile than    
the protostellar cores and that the column densities in Perseus  
are an order of magnitude higher than both the model and 
the Taurus cores.}   
\label{radial}   
\end{figure*}   
   
One of the clear signposts of star formation is direct detection of infall.   
Detection of blue infall profiles 
(see previous chapter by {\it Di Francesco    
et al.}) in optically thick line tracers    
such as HCO$^+$(3--2) toward a number of starless cores in cluster-forming   
regions    
(e.g., OphE-MM2 in $\rho$~Oph -- {\it Belloche et al.}, 2001) suggests that   
some of them are in fact already collapsing and     
on the verge of forming protostars.   
   
Prestellar cores in low-mass proto-clusters also appear to be    
characterized by small core-core relative motions 
(e.g., {\it Walsh et al.}, 2004). For instance,  
based on the    
observed distribution of N$_2$H$^+$(1--0) line-of-sight velocities,   
a global, one-dimensional velocity dispersion    
$\sigma_{1D}$ of $ < 0.4$ km~s$^{-1}$ was found for the cores of the    
$\rho$~Oph cluster ({\it Belloche et al.}, 2001; {\it Andr\'e et al.}, 
2006).   
 
\vspace{0.3cm}  
 
\bigskip   
\centerline{\textbf{ 7. FROM CORES TO PROTOSTARS}}   
\bigskip   
   
Most of the starless cores and all of the   
prestellar cores that we have been discussing    
are expected to evolve into Class~0 protostars    
({\it Andr\'e et al.}, 1993, 2000) and subsequently    
into Class~I protostars ({\it Lada}, 1987; {\it Wilking   
et al.}, 1989).    
Therefore, another approach to constraining    
the initial conditions for protostellar collapse consists    
of studying the structure of young Class~0 protostars.   
These objects are observed early    
enough after point mass formation that they still retain some memory of    
their initial conditions   
(cf. {\it Andr\'e et al.}, 2000).   
   
By comparing the properties of prestellar cores with those of Class~0 cores,   
we can hope to bracket the physical conditions at point mass formation.   
Furthermore, since Class~0 objects have already begun to form stars   
at their centres, we can be sure that they are participating in the   
star-formation process (which is not certain for all starless cores).   
In fact, in some cases it is difficult to differentiate between the most    
centrally-condensed prestellar cores and the youngest Class~0 protostars.    
Examples of low-luminosity, very young Class~0 protostars that look like   
pre-stellar cores include L1014 ({\it Young et al.}, 2004;    
{\it Bourke et al.}, 2005)    
and MC27/L1521F ({\it Onishi et al.}, 1999) -- see also previous chapter    
by {\it Di Francesco et al.}   
   
When discussing the density and velocity structure of Class~0 envelopes,   
we here again contrast isolated and clustered objects.   
In terms of their density profiles, protostellar envelopes are   
found to be more strongly centrally condensed than prestellar cores,   
and do not exhibit any marked inner flattening in their    
radial column density profiles (see Figure~\ref{radial}),   
unlike prestellar cores ({\it J. Kirk et al.}, 2005).   
   
In regions of isolated star formation such as Taurus,   
protostellar envelopes   
have radial density gradients consistent with    
$\rho(r) \propto r^{-p}$ with $p \sim $~1.5--2    
over $\sim $~10000--15000~AU in radius    
(e.g., {\it Chandler and Richer}, 2000;    
{\it Hogerheijde and Sandell}, 2000;   
{\it Shirley et al.}, 2000, 2003; {\it Motte and Andr\'e}, 2001).    
Furthermore, the absolute level of the density    
distributions observed towards   
Taurus Class~0 sources is roughly consistent with    
the predictions of spontaneous collapse models (see Figure~7) 
starting from quasi-equilibrium, thermally-dominated prestellar cores    
(e.g., {\it Hennebelle et al.}, 2003).   
   
By contrast, in cluster-forming regions such as Serpens, Perseus, or    
$\rho$~Oph, Class~0 envelopes are clearly not scale-free.   
They merge either with other cores or other protostellar envelopes,    
or the ambient cloud, at a finite   
radius $\rout \simlt 5\,000$~AU ({\it Motte et al.}, 1998;     
{\it Looney et al.}, 2003).    
They are also typically an order of magnitude more dense than    
models of the spontaneous collapse of isothermal 
(e.g., Bonnor-Ebert) spheres   
predict immediately after point mass formation    
(cf. {\it Motte and Andr\'e}, 2001) -- see Figure~7. 
 
\begin{figure*}   
\setlength{\unitlength}{1mm}  
\noindent  
\begin{picture}(80,95)  
\put(0,0){\includegraphics{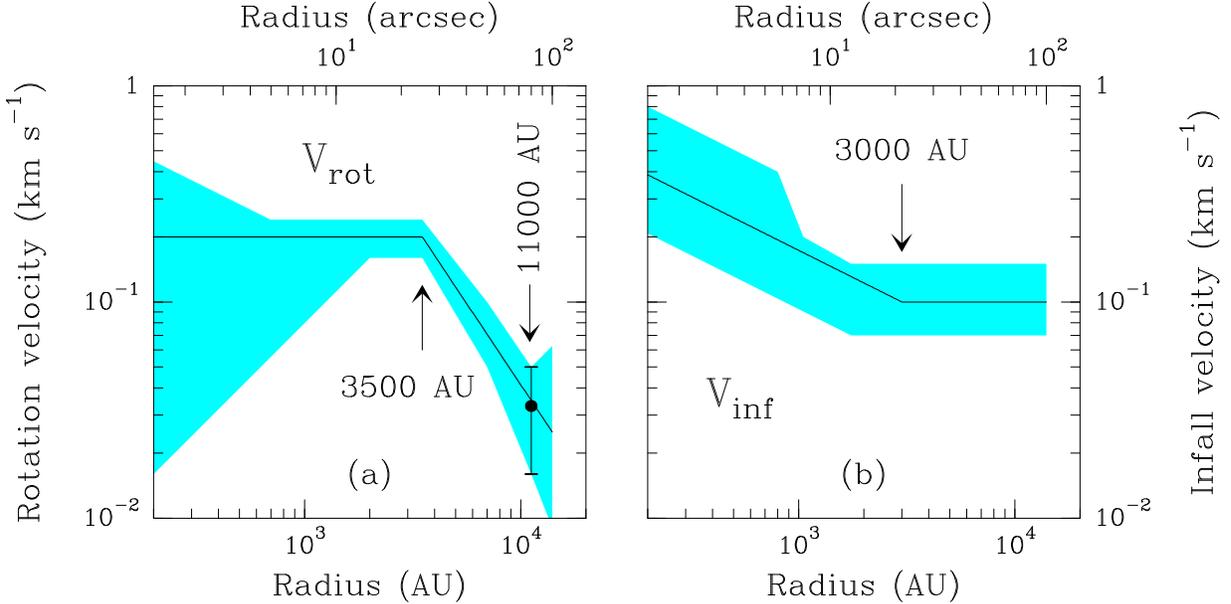}}  
\end{picture}  
\caption{Velocity profiles in the Class~0 protostellar core IRAM~04191.   
Rotational velocity (left) and infall velocity (right) are plotted as   
a function of radius (from {\it Belloche et al.}, 2002). The inner part 
of the envelope is rapidly collapsing and rotating, while the outer 
part undergoes only moderate infall and slower riotation.}   
\label{04191}   
\end{figure*}   
   
Turning to velocity profiles,  
the surrounding environment can play   
an important role in the mass infall rate, 
since in a clustered environment this   
can vary strongly even for protostars with similar final masses   
({\it Klessen}, 2001b). Models suggest that the   
mass infall rate may be a strongly varying function of time, with a   
peak infall rate occurring in the Class 0 stage   
(e.g., {\it Henriksen et al.}, 1997; {\it Whitworth and    
Ward-Thompson}, 2001; {\it Schmeja and Klessen}, 2004). The mean mass   
infall rate is also predicted to decrease as the Mach number of    
the turbulence increases ({\it Schmeja and Klessen}, 2004).    
In addition, the relative   
importance of turbulent and gravitational energy can    
change the number of binary systems that are formed as well as   
their properties, such as semi-major axis and mass ratio ({\it Goodwin   
et al.}, 2004).   
 
There have been few detailed studies of velocity profiles, but 
one example of an isolated Class~0 object is IRAM~04191    
in Taurus (see Figure~\ref{04191} and {\it Belloche et al.}, 2002).    
In this case,   
the inner part of the envelope ($r \simlt 2000-4000$ AU) is    
rapidly collapsing and rotating, while the outer part    
($4000 \simlt r \simlt 11000$~AU) undergoes only moderate infall and    
slower rotation.   
This dramatic drop in rotational velocity beyond $r \sim 4000$~AU,    
combined with the flat infall velocity profile,    
suggests that angular momentum is conserved    
in the collapsing inner envelope but efficiently dissipated,   
perhaps due to     
magnetic braking, in the slowly contracting outer envelope.   
   
The mass infall rate of IRAM~04191 is estimated to be    
$\dot{M}_{inf} \sim 3 \times 10^{-6}$~M$_\odot$~yr$^{-1}$,    
which is $\sim 2-3$ times the canonical $ a_s^3/G $ value    
often used (where $a_s \sim 0.15-0.2\, \kms $    
is the isothermal sound speed). Similar $\dot{M}_{inf}$ values have    
been reported for several other bona-fide Class~0 and Class~I protostars    
in Taurus (e.g., {\it Ohashi}, 1999; {\it Hirano et al.}, 2002).   
   
A very different example in a clustered region   
is IRAS~4A in the NGC~1333 protocluster.   
{\it Di Francesco et al.} (2001) observed inverse P-Cygni    
profiles towards IRAS~4A, from    
which they derived a very large mass infall rate    
$\sim 1.1 \times 10^{-4}$~M$_\odot$~yr$^{-1}$    
at $r \sim 2000$~AU. A similar infall rate was independently    
found by {\it Maret et al.} (2002).    
   
This value of   
$\dot{M}_{inf}$ corresponds to more than $\sim 15$ times    
the canonical $a_{eff}^3/G$ value (where    
$a_{eff} \simlt 0.3\, \kms$ is the effective sound speed).     
This high infall rate results both from a very dense    
envelope and a large, supersonic infall    
velocity -- $\sim 0.68\, \kms $ at $\sim 2000$~AU    
({\it Di Francesco et al.}, 2001).    
   
Other examples of Class~0 protostars in cluster-forming regions with    
quantitative estimates of the mass infall rate include    
NGC~1333-IRAS2, Serpens-SMM4, and IRAS~16293. In all of these    
objects, high $\dot{M}_{inf}$ values     
$ \simgt 3 \times 10^{-5} $~M$_\odot$~yr$^{-1}$ are found   
(e.g., {\it Ceccarelli et al.}, 2000;    
{\it Ward-Thompson and Buckley}, 2001).   
   
The velocity structures of prestellar cores have also been studied in some   
cases. The isolated prestellar core L1544 has been seen to have a `flat'   
velocity profile over a wide range of radii, with no evidence for velocity    
increasing towards the centre    
({\it Tafalla et al.}, 1998; {\it Williams et al.}, 1999).   
Infall profiles have also been observed in a number of other prestellar    
cores at large radii    
(e.g., {\it Lee et al.}, 1999; {\it Gregersen and Evans}, 2000) and    
it seems that a significant number may already be contracting 
(see previous chapter by {\it Di Francesco et al.}). 
   
The observational constraints summarized above have    
strong implications for collapse models.    
First, the extended infall velocity profiles observed in    
prestellar cores and young Class~0 objects are inconsistent    
with pure inside-out collapse and in better agreement with    
isothermal collapse models starting from Bonnor-Ebert spheres    
(e.g., {\it Whitworth and Summers}, 1985; 
{\it Foster and Chevalier}, 1993), or     
similar density profiles (e.g., {\it Whitworth and Ward-Thompson}, 2001).   
   
For isolated cores,    
the fact that the measured infall velocities are subsonic and that there    
is indirect evidence of magnetic braking  
({\it Belloche et al.}, 2002 -- see above) 
suggests that the collapse is     
spontaneous and moderated by magnetic effects in mildly    
magnetized versions of Bonnor-Ebert cloudlets    
(cf. {\it Basu and Mouschovias}, 1994).    
In Taurus, the measured infall rates   
seem to rule out models based on competitive accretion    
(e.g., {\it Bonnell et al.}, 2001) or gravo-turbulent    
fragmentation (e.g. {\it Schmeja and Klessen} 2004)    
which predict large time and spatial variations of~$\dot{M}_{inf}$.   
   
By contrast,    
in protoclusters such as NGC~1333 or $\rho$~Oph, the large overdensity    
factors measured in Class~0 envelopes compared to hydrostatic isothermal    
structures, as well as the supersonic   
infall velocities and very high infall rates observed in some cases,   
are inconsistent with    
self-initiated forms of collapse and require strong external compression.   
 
This point is supported by recent numerical simulations of the collapse 
of Bonnor-Ebert spheres ({\it Hennebelle et al.}, 2003, 2004),    
which show that large overdensity factors,   
together with supersonic infall velocities,    
and high infall rates ($\simgt 10\, \as^3/G$)    
are produced near point mass formation when, and only when, the collapse is    
induced by a strong and rapid increase in external pressure    
(see also {\it Motoyama and Yoshida}, 2003).    
   
The high infall rates at the    
Class~0 stage, as well as the strong decline of $\dot{M}_{inf} $ observed    
between the Class~0 and the Class~I stage in clusters  
(e.g., {\it Henriksen et    
al.}, 1997; {\it Whitworth and Ward-Thompson}, 2001),    
can also be reproduced in the context of the turbulent fragmentation    
picture     
(cf. {\it Schmeja and Klessen}, 2004), according to which dense cores form     
by strong turbulent compression (e.g., {\it Padoan and Nordlund}, 2002).   
   
\bigskip   
\centerline{\textbf{ 8. DISCUSSION AND CONCLUSIONS}}   
\bigskip   
   
We have presented observational results that bear on the evolution of   
dense low-mass cores in an endeavour to estimate which aspects of the   
continuum of models discussed in Section~2 above relate to the different   
environments of star formation that we observe.   
The formation and evolution of cores is crucial to an understanding 
of the star formation process, not least because 
the results presented in Section~5 indicate that the core mass 
function has a very strong bearing on the stellar IMF. 
The results summarized in Section~6 help to discriminate between    
possible theoretical scenarios for the formation and evolution    
of isolated cores compared to cluster-forming cores.   
   
The narrow linewidths observed in prestellar cores in cluster-forming   
regions are in qualitative agreement with the picture according to which such  
cores form by dissipation of internal MHD turbulence  
(e.g., {\it Nakano}, 1998 -- cf. Figure~6).  
These cores may correspond to self-gravitating `kernels',     
of size comparable to the cutoff wavelength for MHD waves    
(e.g., {\it Kulsrud and Pearce}, 1969), that can develop only in    
turbulent cloud cores (e.g., {\it Myers}, 1998).   
   
However, at variance with this picture, we see that   
some cluster-forming cores such as $\rho$~Oph~E  
({\it Belloche et al.}, 2001)    
also exhibit narrow line widths (see Figure~6),  
similar to those of the prestellar cores within them. This tends to 
suggest that spontaneous dissipation of internal MHD turbulence     
may not be the only mechanism responsible for core fragmentation.      
In an alternative view, the formation of cluster-forming cores may   
primarily reflect the action of a strong external trigger at the head of    
elongated, head-tail cloud structures      
(e.g., {\it Tachihara et al.}, 2002; {\it Nutter et al.}, 2006).    
   
A marked increase in external pressure resulting from     
the propagation of neighbouring stellar winds and/or supernova shells    
into a cloud can indeed significantly reduce the critical Bonnor-Ebert mass    
and the corresponding Jeans fragmentation lengthscale    
(cf. {\it Nakano}, 1998).    
It may also trigger protostellar collapse  
(e.g., {\it Boss}, 1995) and account    
for the enhanced infall rates observed at the Class~0 stage in    
cluster-forming clouds (see Section~7).   
   
Furthermore, the small velocity dispersions measured    
for the prestellar cores in the $\rho$~Oph cluster,    
imply a crossing time, $\sim 2 \times 10^6$~yr   
({\it Belloche et al.}, 2001; {\it Andr\'e et al.}, 2006),  
that is larger than the estimated core   
lifetime ($< 2.5 \times 10^5$~yr -- see Section~3). This suggests that    
typical prestellar cores in clusters do not have time to interact    
with one another before collapsing to protostars.    
Taken at face value, this seems inconsistent with models which    
resort to dynamical interactions     
to build up a mass spectrum comparable to the IMF    
(e.g., {\it Bate et al.}, 2003; {\it Bonnell et al.}, 2003).    
Nonetheless these models   
may still be relevant in higher-mass star-forming regions 
(cf. {\it Peretto et al.}, 2006).   
 
Therefore, it appears that the influence of the external 
environment plays a crucial role in the formation and 
evolution of low-mass dense cores. An isolated,   
low-density, quiescent environment will most likely lead to a more   
quasi-static evolution. A clustered, dense environment in which the external   
pressure is increased by the action of nearby, newly-formed stars, will   
probably yield a more dynamic evolutionary scenario.  
 
The fact that most isolated prestellar cores appear to 
be within a factor of a few of magnetic criticality 
suggests that the magnetic field is playing an important 
role and is consistent with the ambipolar diffusion picture 
(see Section 4). 
However, whether or not this role is dominant depends on 
the balance between   
the field strength and the other environmental factors.   
   
\bigskip   
\centerline{\textbf{ 9. FUTURE DEVELOPMENTS}}   
\bigskip   
  
There are many exciting developments in telescopes and instrumentation  
planned in the next few years that will impact this field. These  
include new, more sensitive cameras for single-dish telescopes, 
such as SCUBA-2 on the James Clerk Maxwell Telescope
(JCMT) as well as SPIRE and PACS on the  
Herschel Space Observatory, and new interferometers such as 
the Combined Array for Research in Millimeter-wave  
Astronomy (CARMA) and the Atacama Large Millimetre Array (ALMA). 
 
The new submillimetre cameras on JCMT and Herschel 
will increase our wide-area mapping coverage, so that  
for example, SCUBA-2 and SPIRE 
will map almost all star-forming regions within 0.5\,kpc. 
These observations will produce a flux-limited,  
multi-wavelength snapshot of star formation near the Sun, providing a  
legacy of images, as well as point-source and extended-source  
catalogues covering up to 700 square degrees of sky.  
 
On small scales, the Herschel observations will, for the first 
time, resolve the detailed dust temperature structure of the nearest  
isolated 
prestellar cores. On global molecular cloud scales, the large spatial 
dynamic range of the Herschel images will provide a unique view of the 
formation of both isolated prestellar cores and cluster-forming cores. 
 
CARMA will bring improved angular resolution ($\sim 0.13$ arcsec)  
and sensitivity -- $\sim 25$ times better than
the Berkeley Illinois Maryland Array (BIMA) -- to mm-wave polarization  
studies of the dust and molecular line emission in dense clouds and  
lead to routine high-resolution polarization mapping.   
These instrumental gains will enable Zeeman  
mapping of the CN J=(1$\rightarrow$0)  
transition and measurement of line-of-sight  
magnetic field  
strengths at densities $n(H_2) \sim 10^{5-6}$ cm$^{-3}$ and mapping  
of dust and CO  
linearly polarized emission toward both high-mass and low-mass   
star-formation regions.  
 
The Atacama Large Millimeter Array (ALMA) will have more than an order of  
magnitude greater 
sensitivity and resolution compared to existing millimetre  
arrays. Observations with ALMA will allow us to probe the  
nearby star forming regions discussed in this chapter on spatial scales of  
a few tens of AU, while allowing these types of analyses to be extended to  
the more distant regions of high mass star formation.  
 
ALMA's broad  
wavelength coverage and flexible spectrometer will allow detailed  
studies of cores throughout the submillimetre 
windows, while its dual polarization receivers will allow sensitive high  
resolution observations of magnetic field signatures, both with  
polarization and with Zeeman observations.  
 
Perhaps these instrumental advances will have helped to answer 
some of the questions raised in this chapter in time for PPVI. 
 
\bigskip 
   
{\bf Acknowledgments.} This work was carried out while DWT was on sabbatical  
at the Observatoire de Bordeaux and CEA Saclay, and he would like to thank  
both institutions for their hospitality. RMC received partial research  
support under grants NSF AST 02-28953 and NSF AST 02-05810.  
The research of DJ is supported through a grant from the Natural 
Sciences and Engineering Council of Canada, held at the University of 
Victoria. This work was partially supported by a Discovery 
grant to CW from the Natural Sciences and Engineering  
Research Council of Canada. 
Helen Kirk, David Nutter and Mike Reid are thanked for assistance 
in producing some of the figures for this chapter. 
The referee is thanked for helpful comments on the manuscript. 
   
\bigskip   
   
\centerline\textbf{ REFERENCES} 
\bigskip 
\parskip=0pt 
{\small 
\baselineskip=11pt 
 
\refs 
Aikawa Y., Herbst E., Roberts H., and Caselli P. (2005) 
{\it Astrophys. J., 620}, 330-346. 
 
\refs  
Andr\'e P., Ward-Thompson D., and Barsony M. (1993)  
{\it Astrophys. J., 406}, 122-141.  
 
\refs  
Andr\'e P., Ward-Thompson D., and Barsony M. (2000) In  
{\it Protostars and Planets IV} 
(V. Mannings, A. P. Boss, and S. S. Russell, eds.), pp. 59-96.  
Univ. of Arizona, Tucson.  
 
\refs 
Andr\'e P., Belloche A., Motte F., and Peretto N. (2006)  
{\it Astron. Astrophys.}, in press. 
 
\refs  
Aoyama H., Mizuno N., Yamamoto H., Onishi T., Mizuno A., and Fukui Y. (2001)  
{\it Publ. Astron. Soc. Japan, 53,} 1053-1062. 
 
\refs 
Aso Y., Tatematsu K., Sekimoto Y., Nakano T., Umemoto T., et al. (2000)  
{\it Astrophys. J. Suppl., 131}, 465-482. 
 
\refs 
Ballesteros-Paredes J., Klessen R. S., and Vazquez-Semadeni E. (2003) 
{\it Astrophys. J., 592}, 188-202. 
 
\refs  
Basu S. (2000) {\it Astrophys. J., 540}, L103-106. 
 
\refs  
Basu S. and Mouschovias T. Ch. (1994)  
{\it Astrophys. J., 432}, 720-741.  
 
\refs 
Bate M. R., Bonnell I. A., and Bromm V. (2003)  
{\it Mon. Not. R. Astron. Soc., 339}, 577-599. 
 
\refs 
Beichman C. A., Myers P. C., Emerson J. P., Harris S., Mathieu R.,  
Benson P. J., Jennings R. E. (1986) 
{\it Astrophys. J., 307}, 337-349. 
 
\refs  
Belloche A., Andr\'e P., and Motte F. (2001) In {\it From Darkness to  
Light} (T. Montmerle and P. Andr\'e, eds.), pp. 313-318, ASP, San Francisco. 
 
\refs 
Belloche A., Andr\'e P., Despois D., and Blinder S. (2002)  
{\it Astron. Astrophys., 393}, 927-947. 
 
\refs 
Benson P. J. and Myers P. C. (1989) 
{\it Astrophys. J. Suppl., 71}, 89-108. 
 
\refs 
Beuther H. and Schilke P. (2004) 
{\it Science, 303}, 1167-1169. 
 
\refs 
Bonnell I. A., Bate M. R., Clarke C. J., and Pringle J. E. (2001) 
{\it Mon. Not. R. Astron. Soc., 323}, 785-794. 
 
\refs 
Bonnell I. A., Bate M. R., and Vine S. G. (2003)  
{\it Mon. Not. R. Astron. Soc., 343}, 413-418. 
 
\refs 
Bontemps S., Andr\'e P., Kaas A. A., Nordh L., Olofsson G., et al. (2001) 
{\it Astron. Astrophys., 372}, 173-194. 
 
\refs 
Boss A. P. (1995) {\it Astrophys. J., 439}, 224-236. 
 
\refs 
Bourke T. L., Hyland A. R., and Robinson G. (1995a) 
{\it Mon. Not. R. Astron. Soc., 276}, 1052-1066. 
 
\refs 
Bourke T. L., Hyland A. R., Robinson G., James S. D., and Wright C. M. (1995b) 
{\it Mon. Not. R. Astron. Soc., 276}, 1067-1084. 
 
\refs 
Bourke T. L., Crapsi A., Myers P. C., Evans N. J., Wilner D. J., et al. 
(2005) {\it Astrophys. J., 633}, L129-132. 
 
\refs  
Caselli P., Benson P. J., Myers P. C., and Tafalla M. (2002)  
{\it Astrophys. J., 572}, 238-263. 
 
\refs 
Ceccarelli C., Castets A., Caux E., Hollenbach D., Loinard L., et al. (2000)  
{\it Astron. Astrophys., 355}, 1129-1137. 
 
\refs  
Chabrier G. (2003)  
{\it Publ. Astron. Soc. Pac., 115}, 763-795.    
 
\refs 
Chandler C. J. and Richer, J. S. (2000)  
{\it Astrophys. J., 530}, 851-866. 
 
\refs  
Chandrasekhar S. and Fermi E. (1953)  
{\it Astrophys. J., 118}, 113-115. 
 
\refs 
Ciolek G. E. and Basu S. (2001) 
{\it Astrophys. J., 547}, 272-279. 
 
\refs 
Clemens D. P. and Barvainis R. (1988) 
{\it Astrophys. J., Suppl., 68}, 257-286. 
 
\refs 
Crutcher R. M. (1999) {\it Astrophys. J., 520}, 706-713.  
 
\refs  
Crutcher R.~M. and Troland T.~H. (2000)  
{\it Astrophys. J., 537}, L139-L142. 
 
\refs  
Crutcher R.~M., Troland T.~H., Goodman A.~A., Heiles C., Kaz\'es I.  
and Myers P.~C. (1993) {\it Astrophys. J., 407}, 175-184. 
 
\refs  
Crutcher R.~M., Nutter D., Ward-Thompson D., and Kirk J.~M. (2004)  
{\it Astrophys. J., 600}, 279-285. 
 
\refs 
Di Francesco J., Myers P. C., Wilner D. J., and Ohashi N. (2001)  
{\it Astrophys. J., 562}, 770-789. 
 
\refs 
Elmegreen B. G. (2000) {\it Astrophys. J., 530}, 277-281. 
 
\refs 
Enoch M. L., Young K. E., Glenn J., Evans N. J., Golwala S., et al. (2006) 
{\it Astrophys. J.}, in press. 
 
\refs 
Foster P. N. and Chevalier R. A. (1993)  
{\it Astrophys. J., 416}, 303-311. 
 
\refs  
Gammie C. F., Lin Y., Stone J. M., and Ostriker E. C. (2003)   
{\it Astrophys. J., 592}, 203-216.    
 
\refs 
Goldsmith P. F. and Li D. (2005)  
{\it Astrophys. J., 622}, 938-958. 
   
\refs  
Goodwin S. P., Whitworth A. P., and Ward-Thompson D. (2004)   
{\it Astron. Astrophys., 423}, 169-182.   
   
\refs 
Greene T. P., Wilking B. A., Andr\'e P., Young E. T., and Lada C. J. (1994) 
{\it Astrophys. J., 434}, 614-626. 
 
\refs 
Gregersen E. M. and Evans N. J. (2000) 
{\it Astrophys. J., 538}, 260-267. 
 
\refs  
Hartmann L., Ballesteros-Paredes J., and Bergin E. A. (2001) 
{\it Astrophys. J., 562}, 852-868. 
 
\refs 
Heiles C. and Crutcher R. M. (2005) In {\it Cosmic Magnetic Fields,  
LNP 664}, (R. Wielebinski and R. Beck, eds.) pp. 137-183. 
Springer, Heidelberg. 
 
\refs 
Hennebelle P., Whitworth A. P., Gladwin P. P., and Andr\'e P. (2003)  
{\it Mon. Not. R. Astron. Soc., 340}, 870-882. 
 
\refs 
Hennebelle P., Whitworth A. P., Cha S., and Goodwin S., (2004)  
{\it Mon. Not. R. Astron. Soc., 348}, 687-701. 
 
\refs 
Henriksen R., Andr\'e P., and Bontemps S. (1997) 
{\it Astron. Astrophys., 323}, 549-565. 
 
\refs 
Hirano N., Ohashi N., and Dobashi K. (2002) In 
{\it 8th Asian-Pacific Regional Meeting}  
 (S.~Ikeuchi, J.~Hearnshaw, and T.~Hanawa, eds.), pp. 141-142.  
Ast. Soc. Japan, Tokyo. 
 
\refs 
Hogerheijde M. R. and Sandell G. (2000)  
{\it Astrophys. J., 534}, 880-893. 
 
\refs  
Holland W. S., Robson E. I., Gear W. K., Cunningham C. R., Lightfoot J. F., 
et al. (1999) {\it Mon. Not. R. Astron. Soc., 303}, 659-672.   
   
\refs 
Jessop N. E. and Ward-Thompson D. (2000) 
{\it Mon. Not. R. Astron. Soc., 311}, 63-74. 
 
\refs 
Jijina J., Myers P. C., and Adams F. C. (1999)  
{\it Astrophys. J., Suppl., 125}, 161-236. 
   
\refs  
Johnstone D., Wilson C. D. Moriarty-Schieven   
G., Joncas G., Smith G., Gregersen E., and Fich, M. (2000)  
{\it Astrophys. J., 545}, 327-339.    
  
\refs  
Johnstone D., Fich M., Mitchell G. F., and Moriarty-Schieven 
G. (2001) {\it Astrophys. J., 559}, 307-317.   
  
\refs  
Johnstone D., Di Francesco J., and Kirk H. (2004)  
{\it Astrophys. J., 611}, L45-48. 
 
\refs 
Johnstone D., Matthews H., and Mitchell G. F. (2006) Astrophys. J., in press.  
   
\refs 
Kaas A. A., Olofsson G., Bontemps S., Andr\'e P., Nordh L., et al. (2004)  
{\it Astron. Astrophys., 421}, 623-642.  
 
\refs 
Kandori R., Nakajima Y., Tamura M., Tatematsu K., Aikawa Y., 
{\it et al.} (2005) {\it Astron. J., 130}, 2166-2184. 
 
\refs 
Kenyon S. J. and Hartmann L. (1995)  
{\it Astrophys. J. Suppl., 101}, 117-171. 
 
\refs 
Kirk H. (2005) {\it MSc thesis}, University of Victoria, Canada. 
  
\refs 
Kirk H., Johnstone D., and Di Francesco J. (2006) 
{\it Astrophys. J.}, in press 
   
\refs 
Kirk J. M., Ward-Thompson D., and Andr\'e P. (2005) 
{\it Mon. Not. R. Astron. Soc., 360}, 1506-1526. 
 
\refs 
Kirk J. M., Ward-Thompson D., and Crutcher R. M. (2006)  
{\it Mon. Not. R. Astron. Soc.}, in press. 
 
\refs  
Klessen R. S. (2001a) {\it Astrophys. J., 550}, L77-80.   
    
\refs  
Klessen R. S. (2001b) {\it Astrophys. J., 556}, 837-846.   
   
\refs  
Klessen R. S. and Burkert A. (2000)  
{\it Astrophys. J. Suppl., 128}, 287-319.    
   
\refs  
Klessen R. S., Burkert A., and Bate M. R. (1998)    
{\it Astrophys. J., 501}, L205-208.   
   
\refs  
Kreysa E., Gemuend H. P., Gromke J., Haslam C. G., Reichertz L.,  
et al. (1999) {\it Soc. Phot. Inst. Eng., 3357}, 319-325.     
   
\refs  
Kroupa P. (2002) {\it Science, 295}, 82-91.    
   
\refs 
Kulsrud R. and Pearce W. P. (1969)  
{\it Astrophys. J., 156}, 445-469. 
 
\refs 
Lada C. (1987) In {\it Star Forming Regions}  (M. Peimbert and J. 
Jugaku, eds.), pp. 1-17. Reidel, Dordrecht. 
 
\refs 
Larson R. B. (1981)  
{\it Mon. Not. R. Astron. Soc., 194}, 809-826. 
 
\refs 
Lee C. W. and Myers P. C. (1999)  
{\it Astrophys. J. Suppl., 123}, 233-250. 
 
\refs 
Lee C. W., Myers P. C., and Tafalla M. (1999)  
{\it Astrophys. J., 526}, 788-805. 
 
\refs 
Looney L. W., Mundy L. G., and Welch W. J. (2003)  
{\it Astrophys. J., 592}, 255-265. 
 
\refs 
Loren R. B. (1989) 
{\it Astrophys. J., 338}, 902-924. 
 
\refs  
MacLow M.-M. and Klessen R.~S. (2004)  
{\it Rev. Mod. Phys., 76}, 125-194. 
 
\refs  
MacLow M.-M., Klessen R.~S., Burkert A., and Smith M.~D. (1998)  
{\it Phys. Rev. Lett., 80}, 2754-2757. 
 
\refs   
Maret S., Ceccarelli C., Caux E., Tielens A. G. G. M., and Castets A.  
(2002) {\it Astron. Astrophys., 395}, 573-585. 
 
\refs 
Matthews B. C. and Wilson C. D. (2002)  
{\it Astrophys. J, 574}, 822-833. 
 
\refs 
Matthews B. C., Wilson C. D., and Fiege J. D. (2001)  
{\it Astrophys. J, 562}, 400-423. 
 
\refs 
McKee C. F. (1989) {\it Astrophys. J, 345}, 782-801. 
 
\refs 
McKee C. F. and Tan J. (2003)  
{\it Astrophys. J., 585}, 850-871. 
 
\refs  
Mookerjea B., Kramer C., Nielbock M., and Nyman L. A. (2004)   
{\it Astron. Astrophys., 426}, 119-129.    
   
\refs 
Motoyama K. and Yoshida T. (2003)   
{\it Mon. Not. R. Astron. Soc., 344}, 461-467. 
 
\refs   
Motte F. and Andr\'e P. (2001)  
{\it Astron. Astrophys., 365}, 440-464. 
 
\refs  
Motte F., Andr\'e P., and Neri R. (1998) 
{\it Astron. Astrophys., 336}, 150-172.    
   
\refs  
Motte F., Andr\'e P., Ward-Thompson D., and Bontemps S. (2001)  
{\it Astron. Astrophys., 372}, L41-44.    
   
\refs  
Motte F., Schilke P., and Lis D. C. (2003)  
{\it Astrophys. J., 582}, 277-291. 
 
\refs  
Mouschovias T. Ch. (1991)  
{\it Astrophys. J., 373}, 169-186. 
 
\refs  
Mouschovias T.~Ch. and Ciolek G.~E. (1999) In  
{\it The Origin of Stars and Planetary Systems}  
(C.~J. Lada and N.~D. Kylafis, eds.), pp. 305-339. Kluwer, Dordrecht. 
 
\refs 
Mouschovias T. Ch., Tassis K., and Kunz M. W. (2006)  
{\it Astrophys. J.}, in press. 
 
\refs 
Myers P. C. (1998) {\it Astrophys. J., 496}, L109-112. 
 
\refs  
Myers P. C. (2000) {\it Astrophys. J., 530}, L119-122.   
   
\refs 
Myers P. C., Linke R. A., and Benson P. J. (1983)  
{\it Astrophys. J, 264}, 517-537. 
 
\refs 
Nakano T. (1998) {\it Astrophys. J., 494}, 587-604. 
 
\refs 
Nutter D. J., 2004, {\it PhD Thesis}, Cardiff University. 
 
\refs 
Nutter D. J., Ward-Thompson D., and Andr\'e P. (2006)  
{\it Mon. Not. R. Astron. Soc.,} in press. 
 
\refs 
Ohashi N. (1999) In {\it Star Formation 1999} (T.~Nakamoto, ed.), 
pp. 129-135.  Nobeyama Radio Obser., Nobeyama. 
 
\refs 
Onishi T., Mizuno A., Kawamura A., Ogawa H., and Fukui Y. (1998) 
{\it Astrophys. J, 502}, 296-314. 
 
\refs 
Onishi T., Kawamura A., Abe R., Yamaguchi N., Saito H., et al. (1999)  
{\it Publ. Astron. Soc. Japan, } 51, 871-881. 
 
\refs 
Onishi T., Mizuno A., Kawamura A., Tachihara K., and Fukui Y. (2002)  
{\it Astrophys. J., 575}, 950-973. 
 
\refs 
Ostriker E. C., Gammie C. F., and Stone J. M. (1999)  
{\it Astrophys. J, 513}, 259-274. 
 
\refs  
Padoan P. and Nordlund A. (2002)  
{\it Astrophys. J., 576}, 870-879.    
   
\refs  
Parker E.~N. (1966)  
{\it Astrophys. J., 145}, 811-833. 
 
\refs 
Peretto N., Andr\'e P., and Belloche A. (2006)  
{\it Astron. Astrophys., 445}, 979-998. 
  
\refs  
Reid M. A. (2005) {\it Ph.D. thesis,} McMaster University, Canada.    
   
\refs  
Reid M. A. and Wilson C. D. (2005)  
{\it Astrophys. J., 625}, 891-905.   
   
\refs  
Reid M. A. and Wilson C. D. (2006)  
{\it Astrophys. J.,} in press. 
   
\refs  
Salpeter E. E. (1955) 
{\it Astrophys. J., 121}, 161-167.   
   
\refs  
Schmeja S. and Klessen R. S. (2004)  
{\it Astron. Astrophys., 419}, 405-417.   
   
\refs 
Shirley Y. L., Evans N. J., Rawlings J. M. C., and Gregersen E. M.  
(2000) {\it Astrophys. J., Suppl., 131}, 249-271. 
 
\refs  
Shirley Y. L., Evans N. J., Young K. E., Knez C., and Jaffe D. T.   
(2003) {\it Astrophys. J. Suppl., 149}, 375-403.   
   
\refs 
Shu F. H., Adams F. C., and Lizano S. (1987)  
{\it  Ann. Rev. Astron. Astrophys., 25}, 23-81. 
 
\refs 
Stanke T., Smith M. D., Gredel R., and Khanzadyan T. (2006) 
{\it Astron. Astrophys.}, in press. 
 
\refs 
Tachihara K., Mizuno A., and Fukui Y. (2000)  
{\it Astrophys. J., 528}, 817-840. 
 
\refs 
Tachihara K., Onishi T., Mizuno A., and Fukui Y. (2002)  
{\it Astron. Astrophys., 385}, 909-920. 
 
\refs  
Tafalla M., Mardones D., Myers P.~C., Caselli P., Bachiller R.  
and  Benson P.~J. (1998)  
{\it Astrophys. J., 504}, 900-914. 
 
\refs  
Tassis K. and Mouschovias T.~Ch. (2004)  
{\it Astrophys. J., 616}, 283-287. 
 
\refs  
Tatematsu K., Umemoto T., Kandori R., and Sekimoto Y. (2004)  
{\it Astrophys. J., 606}, 333-340. 
 
\refs  
Testi L. and Sargent A. I. (1998) 
 {\it Astrophys. J., 508}, L91-94.    
   
\refs  
Tilley D. and Pudritz R. E. (2004)  
{\it Mon. Not. R. Astron. Soc., 353}, 769-788.   
   
\refs  
Tothill N. F. H., White G. J., Matthews H. E., McCutcheon   
W. H., McCaughrean M. J., and Kenworthy M. A. (2002)  
{\it Astrophys. J., 580}, 285-304.    
  
\refs 
Vazquez-Semadeni E., Kim J., Shadmehri M., and Ballesteros-Paredes J.  
(2005) {\it Astrophys. J., 618}, 344-359. 
 
\refs 
Walsh A. J., Myers P. C., and Burton M. G. (2004)  
{\it  Astrophys. J, 614}, 194-202. 
  
\refs 
Ward-Thompson D. (2002)  
{\it Science, 295}, 76-81.

\refs 
Ward-Thompson D. and Buckley H. D. (2001)  
{\it Mon. Not. R. Astron. Soc., } 327, 955-983. 
 
\refs 
Ward-Thompson D., Scott P. F., Hills R. E., and Andr\'e P. (1994) 
{\it Mon. Not. R. Astron. Soc., 268}, 276-290. 
 
\refs 
Ward-Thompson D., Motte F., and Andr\'e P. (1999)  
{\it Mon. Not. R. Astron. Soc., 305}, 143-150. 
 
\refs  
Ward-Thompson D., Kirk J.~M., Crutcher R.~M., Greaves J.~S., Holland W.~S.,
and Andr\'e P. (2000)  
{\it Astrophys. J., 537}, L135-L138. 
 
\refs 
Whitworth A. P. and Summers D. (1985)  
{\it Mon. Not. R. Astron. Soc., 214}, 1-25.  
 
\refs 
Whitworth A. P. and Ward-Thompson D. (2001)  
{\it Astrophys. J., 547}, 317-322. 
 
\refs 
Wilking B. A., Lada C. J., and Young E. T. (1989)  
{\it Astrophys. J., 340}, 823-852. 
  
\refs  
Williams J.~P., Myers P.~C., Wilner D.~J., and Di Francesco J. (1999)  
{\it Astrophys. J., 513}, L61-L64. 
 
\refs 
Wolf S., Launhardt R., and Henning T. (2003) 
{\it Astrophys. J., 592}, 233-244. 
 
\refs 
Wood D. O. S., Myers P. C., and Daugherty D. A. (1994)  
{\it Astrophys. J. Suppl., 95}, 457-501.  
 
\refs 
Young C. H., Jorgensen J. K., Shirley Y. L., Kauffmann J., Huard T., 
{\it et al.} (2004) {\it Astrophys. J. Supp., 154}, 396-401. 
 
}   
   
\end{document}